# Extension of activation cross section data of long lived products in deuteron induced nuclear reactions on platinum up to 50 MeV

F. Ditrói[1*], F. Tárkányi[1], S. Takács[1], A. Hermanne[2]

[1]Institute for Nuclear Research, Hungarian Academy of Sciences, Debrecen. Hungary

[2]Cyclotron Laboratory, Vrije Universiteit Brussel, Brussels, Belgium

## Abstract

In the frame of a systematical study of light ion induced nuclear reactions on platinum, activation cross sections for deuteron induced reactions were investigated. Excitation functions were measured in the 20.8 - 49.2 MeV energy range for the $^{nat}$Pt(d,xn)$^{191,192,193,194,195,196m2,196g,198g,199}$Au, $^{nat}$Pt(d,x)$^{188,189,191,195m,197m,197g}$Pt and $^{nat}$Pt(d,x)$^{189,190,192,194m2}$Ir reactions by using the stacked foil irradiation technique. The experimental results are compared with previous results from the literature and with the theoretical predictions in the TENDL-2014 and TENDL-2015 libraries. The applicability of the produced radio-tracers for wear measurements has been presented.

Keywords: platinum targets; stacked foil technique; deuteron induced reactions; theoretical nuclear reaction model codes; Pt, Au and Ir radioisotopes; Thin Layer Activation

* Corresponding author: ditroi@atomki.hu

## 1. Introduction

The excitation functions of deuteron induced nuclear reactions on platinum were measured to get reliable experimental data for different applications using targets with natural isotopic composition or highly enriched Pt isotopes. Experimental results for production of the medically relevant $^{198}$Au and $^{199}$Au radioisotopes (obtained through $^{198}$Pt(d,2n) and $^{198}$Pt(d,n) + $^{198}$Pt(d,p) reactions) were published by us earlier in [1]. These results were completed with excitation functions of deuteron induced reactions on $^{nat}$Pt up to 40 MeV for other activation products [1-5]. We have earlier also investigated proton induced reactions on Platinum up to 70 MeV energy [5]. At the time of our last study on cross section data for $^{nat}$Pt + deuterons no experimental results were found expect our earlier data (only thick target yield for production of $^{199}$Au in [6]). Since than only two reports were published in [7, 8] up to 12 MeV for a few reaction products.

During our systematic investigations on activation cross sections of deuteron induced reactions we recently had possibility to extend the energy range up to 50 MeV deuterons and these results are presented here.

## 2. Experimental

Elemental cross-sections were measured by using the activation method and the standard stacked foil irradiation technique combined with high-resolution gamma-ray spectroscopy.

Thin Pt target foils (49 μm thick, total number 31), HgS targets (4 mg/cm$^2$) sedimented on Al backing (50 μm) and additional Al monitor foils (50 μm) were stacked and irradiated with a 50 MeV primary energy deuteron at the CGR930 cyclotron in Louvain la Neuve (LLN). The irradiation was done in a Faraday cup like target holder provided with a long collimator. The nominal beam current was about 70 nA during the 60 minutes' irradiation.

The activity of the irradiated samples was measured non-destructively, without chemical separation, by using HPGe detectors. The samples were measured four times: the first measurements included every second Pt foil and was performed about 6 hours after EOB. The second series of measurements, on the second half of the samples, was performed after one day. The third series (involving all samples) was measured 3-5 days after EOB while a fourth series was done about one-two months after EOB. The large number of simultaneously irradiated targets, limitations in detector capacity and available acquisition time influenced optimization of measuring time. The sample detector distance varied between 25 and 5 cm depending on the activity of the particular target foil in order to keep the dead time low and to optimize the measuring time.

The large number of monitor foils used allowed us to follow the beam intensity and the energy degradation along the stack simultaneously via the $^{nat}Al(d,x)^{22,24}Na$ monitor reactions. The activation cross sections of the simultaneously irradiated mercury have already been published in [9], where the figure of re-measured excitation function of monitor reactions can be found.

The preliminary median deuteron beam energy in each target foil was determined from the incident bombarding energy, stack composition and foil thickness by calculation [10] and was corrected on the basis of the simultaneously re-measured excitation function of the monitor reactions [11, 12]. The uncertainty of the median beam energy on the first foil was estimated by the accelerator staff to be ±0.3 MeV. The uncertainty for the energy is increasing throughout the stack and reaches a maximum of ±1.2 MeV in the last foil due to possible variations and inhomogeneity in thickness of target foils in the stack, incident energy spread and to straggling effects.

So called elemental cross-sections were determined considering that the Pt target was monoisotopic. Direct and/or cumulative cross-sections were calculated depending on the contributing processes and to the activity measurements. The decay data used are collected in Table 1 and were taken from NuDat 2.0 database [13], while the indicated Q values of the contributing reactions originate from the BNL Q-value calculator [14]. The determination of the cross-sections was iterative: in case of contradictions between different measurements in time, the spectrum unfolding and the data evaluation was repeated taking into account the earlier experimental data, the theoretical predictions, the systematic trends, parent–daughter relationships, etc.

The uncertainties of the cross were estimated in the standard way [15], the following independent relative errors were summed quadratically: determination of the peak areas including statistical errors (2–15%), the number of target nuclei including non-uniformity (5%), detector efficiency (5%) and incident particle intensity (7%). The total uncertainty of the cross-section values was evaluated to be between 10–17%.

**Table 1.** Decay characteristic of the investigated reaction products and Q-values of the contributing processes.

| Nuclide<br>Spin/parity<br>Level energy | Half-life | $E_\gamma$(keV) | $I_\gamma$ (%) | Contributing process | Q-value(MeV) |
|---|---|---|---|---|---|
| $^{191g}Au$<br>$3/2^+$ | 3.18 h | 277.86<br>399.84<br>478.04<br>586.44<br>674.22 | 6.4<br>4.2<br>3.5<br>15.0<br>6.0 | $^{190}Pt(d,n)$<br>$^{192}Pt(d,3n)$<br>$^{194}Pt(d,5n)$<br>$^{195}Pt(d,6n)$<br>$^{196}Pt(d,7n)$<br>$^{198}Pt(d,9n)$ | 1.55<br>-13.56<br>-28.17<br>-34.28<br>-42.20<br>-55.60 |
| $^{192}Au$<br>$1^-$ | 4.94 h | 295.96<br>308.46<br>316.51<br>612.46 | 23<br>3.5<br>59<br>4.4 | $^{192}Pt(d,2n)$<br>$^{194}Pt(d,4n)$<br>$^{195}Pt(d,5n)$<br>$^{196}Pt(d,6n)$<br>$^{198}Pt(d,8n)$ | -6.52<br>-21.14<br>-27.24<br>-35.16<br>-48.56 |
| $^{193g}Au$<br>$3/2^+$ | 17.65 h | 112.52<br>173.52<br>255.57<br>268.22<br>439.04 | 2.2<br>2.7<br>6.2<br>3.6<br>1.78 | $^{192}Pt(d,n)$<br>$^{194}Pt(d,3n)$<br>$^{195}Pt(d,4n)$<br>$^{196}Pt(d,5n)$<br>$^{198}Pt(d,7n)$ | -6.52<br>-12.43<br>-18.54<br>-26.46<br>-39.86 |
| $^{194}Au$<br>$1^-$ | 38.02 h | 293.55<br>328.46<br>1468.88 | 10.58<br>60.4<br>6.61 | $^{194}Pt(d,2n)$<br>$^{195}Pt(d,3n)$<br>$^{196}Pt(d,4n)$<br>$^{198}Pt(d,6n)$ | -5.56<br>-11.66<br>-19.58<br>-32.98 |
| $^{195g}Au$<br>$3/2^+$ | 186.01 d | 98.86<br>129.74 | 11.21<br>0.84 | $^{194}Pt(d,n)$<br>$^{195}Pt(d,2n)$<br>$^{196}Pt(d,3n)$<br>$^{198}Pt(d,5n)$ | 2.87<br>-3.23<br>-11.15<br>-24.56 |
| $^{196m2}Au$<br>$12^-$<br>595.66 keV | 9.6 h | 147.81<br>168.37<br>188.27 | 43.5<br>7.8<br>30.0 | $^{195}Pt(d,n)$<br>$^{196}Pt(d,2n)$<br>$^{198}Pt(d,4n)$ | 3.41<br>-4.51<br>-17.92 |
| $^{196g}Au$<br>$2^-$ | 6.1669 d | 333.03<br>355.73 | 22.9<br>87 | $^{195}Pt(d,n)$<br>$^{196}Pt(d,2n)$<br>$^{198}Pt(d,4n)$ | 3.41<br>-4.51<br>-17.92 |
| $^{198m}Au$<br>$12^-$<br>811.7 keV | 2.272 d | 97.21<br>180.31<br>204.10<br>214.89<br>333.82 | 69<br>49<br>39<br>77.3<br>18 | $^{198}Pt(d,2n)$ | -3.33 |
| $^{198g}Au$<br>$2^-$ | 2.6947 d | 411.80 | 95.62 | $^{198}Pt(d,2n)$ | -3.33 |
| $^{199}Au$<br>$3/2^+$ | 3.139 d | 158.38<br>208.20 | 40.0<br>8.72 | $^{198}Pt(d,n)$ | 42.53 |
| $^{188}Pt$<br>$0^+$ | 10.2 d | 187.59<br>195.05<br>381.43<br>423.34 | 19.4<br>18.6<br>7.5<br>4.4 | $^{190}Pt(d,p3n)$<br>$^{192}Pt(d,p5n)$<br>$^{194}Pt(d,p7n)$<br>$^{195}Pt(d,p8n)$<br>$^{196}Pt(d,p9n)$ | -17.86<br>-32.97<br>-47.59<br>-53.69<br>-61.61 |
| | | | | $^{198}$Pt(d,p11n) | |
| $^{189}$Pt<br>3/2$^{-}$ | 10.87 h | 243.50<br>568.85<br>721.38 | 7.0<br>7.1<br>9.3 | $^{190}$Pt(d,p2n)<br>$^{192}$Pt(d,p4n)<br>$^{194}$Pt(d,p6n)<br>$^{195}$Pt(d,p7n)<br>$^{196}$Pt(d,p8n)<br>$^{198}$Pt(d,p10n) | -11.14<br>-26.24<br>-40.86<br>-46.96<br>-54.89 |
| $^{191}$Pt<br>3/2$^{-}$ | 2.802 d | 96.517<br>129.42<br>172.18<br>351.21<br>359.90<br>409.44<br>538.90 | 3.28<br>3.2<br>3.52<br>3.36<br>6.0<br>8.0<br>13.7 | $^{190}$Pt(d,p)<br>$^{192}$Pt(d,p2n)<br>$^{194}$Pt(d,p4n)<br>$^{195}$Pt(d,p5n)<br>$^{196}$Pt(d,p6n)<br>$^{198}$Pt(d,p8n) | 4.22<br>-10.89<br>-25.50<br>-31.60<br>-39.53<br>-52.93 |
| $^{193m}$Pt<br>13/2$^{+}$<br>149.78 keV | 4.33 d | 135.50 | 0.112 | $^{192}$Pt(d,p)<br>$^{194}$Pt(d,p2n)<br>$^{195}$Pt(d,p3n)<br>$^{196}$Pt(d,p4n)<br>$^{198}$Pt(d,p6n) | 4.04<br>-10.58<br>-16.68<br>-24.60<br>-38.01 |
| $^{195m}$Pt<br>13/2$^{+}$<br>259.29 keV | 4.010 d | 98.90<br>129.79 | 11.7<br>2.90 | $^{194}$Pt(d,p)<br>$^{195}$Pt(d,pn)<br>$^{196}$Pt(d,p2n)<br>$^{198}$Pt(d,p4n) | 3.88<br>-22.25<br>-10.15<br>-23.55 |
| $^{197m}$Pt<br>13/2$^{+}$<br>399.59 keV | 95.41 min | 346.5 | 11.1 | $^{196}$Pt(d,p)<br>$^{198}$Pt(d,p2n) | 3.62<br>-9.78 |
| $^{197g}$Pt<br>1/2$^{+}$ | 19.89 h | 191.44 | 3.7 | $^{196}$Pt(d,p)<br>$^{198}$Pt(d,p2n) | 3.62<br>-9.78 |
| $^{188}$Ir<br>1$^{-}$ | 41.5 h | 155.05<br>477.99<br>633.02 | 30<br>14.7<br>18 | $^{190}$Pt(d,2p2n)<br>$^{192}$Pt(d,2p4n)<br>$^{194}$Pt(d,2p6n)<br>$^{195}$Pt(d,2p7n)<br>$^{196}$Pt(d,2p8n)<br>$^{198}$Pt(d,2p10n) | -16.56<br>-31.67<br>-46.28<br>-52.39<br>-60.31 |
| $^{189}$Ir<br>3/2$^{+}$ | 13.2 d | 245.08 | 6.0 | $^{190}$Pt(d,2pn)<br>$^{192}$Pt(d,2p3n)<br>$^{194}$Pt(d,2p5n)<br>$^{195}$Pt(d,2p6n)<br>$^{196}$Pt(d,2p7n)<br>$^{198}$Pt(d,2p9n) | -8.38<br>-23.49<br>-38.10<br>-44.21<br>-52.13 |
| $^{190}$Ir<br>4$^{-}$ | 11.78 d | 186.68<br>361.09<br>371.24<br>407.22<br>518.55<br>557.95<br>569.30<br>605.14 | 52<br>13.0<br>22.8<br>28.5<br>34.0<br>30.1<br>28.5<br>39.9 | $^{190}$Pt(d,2p)<br>$^{192}$Pt(d,2p2n)<br>$^{194}$Pt(d,2p4n)<br>$^{195}$Pt(d,2p5n)<br>$^{196}$Pt(d,2p6n)<br>$^{198}$Pt(d,2p8n) | -2.01<br>-17.12<br>-31.73<br>-37.84<br>-45.76<br>-59.16 |
| $^{192g}$Ir<br>4$^{+}$ | 73.829 d | 295.95<br>308.46<br>316.51<br>468.07 | 28.71<br>29.70<br>82.86<br>47.84 | $^{192}$Pt(d,2p)<br>$^{194}$Pt(d,2p2n)<br>$^{195}$Pt(d,2p3n)<br>$^{196}$Pt(d,2p4n) | -2.90<br>-17.51<br>-23.61<br>-31.54 |
| | | 604.41 | 8.216 | $^{198}Pt(d,2p6n)$ | -44.94 |
| $^{194m2}Ir$<br>(10,11)<br>190.0+x keV | 171 d | 328.5<br>338.8<br>390.8<br>482.6<br>562.4<br>600.5<br>687.8 | 93.0<br>55.0<br>35.0<br>97.0<br>35.0<br>62.0<br>59.0 | $^{194}Pt(d,2p)$<br>$^{195}Pt(d,2pn)$<br>$^{196}Pt(d,2p2n)$<br>$^{198}Pt(d,2p4n)$ | -3.67<br>-9.78<br>-17.70<br>-31.10 |

Abundance of isotopes of natural Pt (%): $^{190}Pt$ - 0.01, $^{192}Pt$ - 0.79, $^{194}Pt$ - 32.9, $^{195}Pt$ - 33.8, $^{196}Pt$ -25.3, $^{198}Pt$ – 7.2.
The *Q*-values shown in Table 2 refer to formation of the ground state. Decrease Q-values for isomeric states with level energy of the isomer
pn→d +2.2 MeV, p2n→t +8.5 MeV, 2pn→$^3$He +7.7 MeV, 2p2n→α +28.3 MeV

## 3. Results

The measured experimental cross-section data are shown in Figs 1-19. The numerical values are essential for further evaluation and dissemination and are presented in Tables 2 - 3. The experimental data are also compared with the results given in the on-line libraries TENDL-2014 [16] and TENDL-2015 [17] calculated with the latest version of the TALYS code family [18]. The investigated radio-products are produced directly and/or additionally through the decay of a shorter-lived isobaric parent. The ground state of the produced radioisotopes can, apart from the direct production, also be produced through the internal transition of an isomeric state. Due to the large number of foils and the limited detector capacity only every second foil was measured in the first and the second series of gamma spectra measurement. It resulted cross sections for some shorter-lived radio-products only at every second energy point.

### 3.1 Radioisotopes of gold

The cross section values of the investigated gold radionuclides are due to direct production via (d,xn) reactions, except $^{199}$Au through the $\beta^-$- decay of the shorter lived isobaric parent $^{199}$Pt radioisotope

#### *3.1.1 $^{nat}$Pt(d,n)$^{199}$Au(cum) reaction*

The $^{199}$Au ($T_{1/2}$ = 3.139 d) is produced directly through the $^{198}$Pt(d,n) reaction. The cross-sections are however cumulative as they were measured after the complete $\beta^-$ -decay of the parent $^{199}$Pt ($T_{1/2}$ = 30.8 min) produced through the $^{198}$Pt(d,p) process (Fig. 1). Our new experiment gives an extension of the cross section data for the energy region between 40 and 50 MeV and shows, in spite of some scattered points, an agreement with our previous results [4] as well as connects properly with the recent data from Khandaker et al. [8]. The TENDL calculations describe the trend of the excitation function, but underestimate the experimental values.

### 3.1.2 $^{nat}Pt(d,xn)^{198g}Au$ reaction

Independent cross sections for direct production of the ground state of $^{198}Au$ ($T_{1/2}$ = 2.6947 d) were deduced (Fig. 2). No corrections could be done for the contribution from decay of the $^{198m}Au$ ($T_{1/2}$ = 2.272 d, IT 100 %) due to low cross section value assessed from the lower energy experimental data of the present measurement, and due to the fact that the gamma-lines from the decay of the similar half-life $^{198m}Au$ were present in the gamma-spectra with very low statistics. Our new experimental results are slightly higher than our previous data [4], connect well with the recent data of Khandaker et al. [8] but the connection with the results of Tárkányi et al. [3] is not acceptable.

### 3.1.3 $^{nat}Pt(d,xn)^{196m2}Au$ reaction

Independent cross-sections for the $T_{1/2}$ = 9.6 h isomeric state $^{196m2}Au$ are shown in Fig. 3 in comparison with the earlier results and data in the TENDL libraries. Our new are slightly lower than our previous results in the energy region above 20 MeV but connect well with the previous experimental data of Tárkányi et al. [3]. The overlapping with the recent results of Khandaker et al. [8] is also good. The TENDL estimations describe well the shape (double peak) of the experimental excitation function including the positions of the peaks, but strongly overestimate the values.

### 3.1.4 $^{nat}Pt(d,xn)^{196g}Au(cum)$ reaction

The cross sections for cumulative formation of the ground state $^{196g}Au$ ($T_{1/2}$ = 6.1669 d) were determined after complete decay of the two higher laying isomeric states (m2 - $T_{1/2}$ = 9.6 h, m1 - $T_{1/2}$ = 8.1 s) (Fig. 4). Our new data are higher than our previous data [4] and the connection to the previous results of Tárkányi et al. [3] is acceptable, the recent results of Khandaker et al. [8] are slightly larger in the overlapping energy range. The TENDL approximation describes the shape and the maximum positions acceptable well, but slightly underestimates above 30 MeV and overestimates below this energy.

### 3.1.5 $^{nat}Pt(d,xn)^{195}Au(cum)$ reaction

The cumulative cross-sections were determined after the decay of the $T_{1/2}$ = 30.5 s isomeric state (Fig. 5). In this case the TENDL results give quite a good approximation of the experimental values both for the shape and for the values. The new data are very scattered and in average higher than our previous results [3, 4] but connect well with the recent data of Khandaker et al. [8] in the overlapping energy region.

### 3.1.6 $^{nat}Pt(d,xn)^{194}Au$ reaction

The excitation function for production of $^{194}Au$ ($T_{1/2}$ = 38.02 h) were determined via the 1468 keV gamma-line to avoid and minimize the contribution of common gamma-lines from the decay of long-lived $^{194m}Ir$ ($T_{1/2}$ = 171 d) and $^{194g}Ir$ ($T_{1/2}$ = 19.28 h)(Fig. 6). It should be mentioned from other side that the contribution of $^{194}Ir$ will be small due to the low cross sections (see section 3.3.1). In this case the shape of the TENDL curves is good, including the numerical agreement. The data of Khandaker et al. [8] agree well in the overlapping region and the connection with the results of Tárkányi et al. [3] is good. Our previous data [4] are lower in the whole energy region.

### 3.1.7 $^{nat}Pt(d,xn)^{193}Au(cum)$ reaction

The measured cross-sections of $^{193g}Au$ ($T_{1/2}$ = 17.65 h) include the internal decay of the $T_{1/2}$ = 3.9 s isomeric state (Fig. 7) and in such a way are cumulative. Our new data show partial agreement with our previous measurement [4] and are slightly lower than both other results of us and Khandaker et al. [3, 8]. The seen discrepancy (grouping) in our new data especially in the higher energy range might come from the different measurement series or from target foil inhomogeneity. The TENDL curves overestimate the experimental values again.

### 3.1.8 $^{nat}Pt(d,xn)^{192}Au$ reaction

The activation cross sections for production of $^{192}$Au ($T_{1/2}$ = 4.94 h) are shown in Fig. 8. In this case the agreement with our previous data is acceptable. The connection with the previous data sets [3, 8] is good. The TENDL estimations fit well with the experimental values up to 35 MeV and underestimates above this energy.

#### *3.1.9 $^{nat}$Pt(d,xn)$^{191}$Au reaction*

The cross section data of $^{191}$Au ($T_{1/2}$ = 3.18 h) include the contribution from the decay of the simultaneously produced short half-life isomeric state ($T_{1/2}$ = 0.92 s, 100% IT) (Fig. 9). The only previous data set is also from our group [4], and the previous data points are slightly lower than our new results. The trend and values given by the TENDL calculations are acceptable good.

### 3.2 Radioisotopes of platinum

The isotopes of platinum are produced via (d,pxn) reactions and through the EC-$\beta^+$ decay of Au and $\beta^-$- decay of Ir parent radionuclides.

#### *3.2.1 $^{nat}$Pt(d,X)$^{197m}$Pt reaction*

Due to the long cooling time before the first measurement the data for production of $^{197m}$Pt metastable state ($T_{1/2}$ = 95.41 min) could be assessed with relatively poor statistical significance as shown in Fig. 10. The magnitude and trend of the scattered cross sections are in the range of the theoretical predictions. No previous data were found in the literature.

#### *3.2.2 $^{nat}$Pt(d,X)$^{197g}$Pt(m+) reaction*

The measured production cross-section of $^{197g}$Pt ($T_{1/2}$ = 19.8915 h) is cumulative, including the contribution from internal transition ($T_{1/2}$ = 95.41 min, IT: 96.7) of the isomeric state and from the decay of the short-lived $^{197m,g}$Ir isomers ($T_{1/2}$ = 5.8 min and 8.9 min) (Fig. 11). Our new results give an acceptable connection to the previous data of [3, 8], and they are larger than our previous results [4]. The TENDL curves describe the trend of the experimental excitation function but give lower estimation.

#### 3.2.3 $^{nat}Pt(d,X)^{195m}Pt$ reaction

The $^{195}Pt$ has a high spin longer-lived isomeric state ($T_{1/2}$ = 4.02 d) decaying to the stable ground state. The excitation function of the isomeric state is shown in Fig. 12. Our new data agree well with the previous experimental data in the overlapping energy regions and give good connection to them. The TENDL estimations overestimate the experimental values.

#### 3.2.4 $^{nat}Pt(d,X)^{191}Pt(cum)$ reaction

The cumulative cross-section for production of $^{191}Pt$ ($T_{1/2}$ = 2.802 d) includes direct production and formation through decay of $^{191}Au$ ($T_{1/2}$ = 3.18 h) and is shown in Fig. 13. The targets were measured after the complete decay of the parent isotope, so in such a way the presented results are cumulative. Our new data are in good agreement with our previous results [4], as well as the TENDL-calculations give a good approximation up to 42 MeV but underestimate above this energy.

#### 3.2.5 $^{nat}Pt(d,X)^{189}Pt(cum)$ reaction

The cross-sections for production of $^{189}Pt$ ($T_{1/2}$ = 11 h) is shown in Fig. 14 and were determined after the complete decay of the two longer-lived states of the parent $^{189}Au$ ($T_{1/2}$ = 4.6 min and 28.3 min). Only two points above 45 MeV were found indicating that there is no measurable contribution from low abundance Pt target isotopes (see Q-values of contributing reactions in Table 2). The new measured points are fit on the theoretical model calculation curve(s) acceptable well. No previous experimental data were found. The two TENDL versions give different approximation above 50 MeV, but unfortunately our data cannot support any of them because the difference is seen only above our highest energy point.

### *3.2.6 $^{nat}Pt(d,X)^{188}Pt(cum)$ reaction*

Only a single cross section point for formation of $^{188}$Pt ($T_{1/2}$ = 10.2 d) was obtained (Fig. 15). The cross-section is cumulative (measured after the complete decay of the parent $^{188}$Au) and contains the contribution of the direct production and the simultaneous decay of $^{188}$Au ($T_{1/2}$ = 8.8 min). The measured single point shows a good agreement with the TENDL-2014 prediction. No previous literature data were found for this reaction in the literature.

## 3.3 Radioisotopes of iridium

The radioisotopes of iridium are produced directly by (p,2pxn) reactions or through EC-$\beta$+ decay of simultaneously formed parent Pt radionuclides. For production of the investigated radioisotopes there was no contribution from the simultaneously produced Os radioisotopes.

### *3.3.1 $^{nat}Pt(d,X)^{194m2}Ir$ reaction*

The $^{194}$Ir radioisotope has three states: the $T_{1/2}$ = 19.28 h ground state, the $T_{1/2}$ = 31.35 ms very first short-lived metastable state and the long-lived (171 d) second metastable state. The strongest gamma-line of the ground-state (328.448 keV, 13.1 %) interfere with one of the gamma-lines of $^{194}$Au (328.464 keV, 60.4 %). No other gamma-lines of $^{194g}$Ir could be identified in the spectra. In such a way the measured spectra allow to get cross sections only for the long-lived metastable state. The peak statistics is low, the cross sections have large uncertainties, but the different gamma-lines gave similar results (Fig. 16). Our new data, in spite of the large scattering, show acceptable agreement and good connection with the recent experimental data of Khandaker et al. from the literature [8]. The newest TENDL overestimates the experimental results.

### *3.3.2 $^{nat}Pt(d,X)^{192}Ir(cum)$ reaction*

The $^{192}$Ir radioisotope has also three states. The ground state has $T_{1/2}$ = 3.8 d half-life. The short-lived isomeric state $^{192m1}$Ir ($T_{1/2}$ = 1.45 min) decays for more than 99.9% to the ground state. The second isomeric state has long half-life ($T_{1/2}$ = 241 a) and decays for 100% by internal transition. Our experimental data (Fig. 17) include the direct production of the ground state including the decay of the short-lived isomeric state. The new data are higher than our previous results in the overlapping energy region but agree well with the previous data of Tárkányi et al. and Khandaker et al. [3, 8]. Both versions of TENDL give lower approximations.

### *3.3.3 $^{nat}Pt(d,X)^{190}Ir(m+)$ reaction*

The isotope $^{190}$Ir has three longer-lived states. The ground state ($T_{1/2}$ = 11.8 d), the first metastable state $^{190m1}$Ir ($T_{1/2}$ = 1.2 h), which completely decays to the ground state, and the second high-spin isomeric state $^{190m2}$Ir ($T_{1/2}$ = 3.25 h), which decays only for 5.6 % by internal transition to the ground state. The possible isobaric parents are stable or alpha particle emitters. The experimental cross-sections for production of the ground state contain the decay of the two isomeric states in addition to the direct production (Fig. 18). Our new data are slightly higher than the previous data of Khandaker et al. [8] but the TENDL codes give much lower approximation.

### *3.3.4 $^{nat}Pt(d,X)^{189}Ir(cum)$ reaction*

The determined cross-section of $^{189}$Ir ($T_{1/2}$ = 13.2 d) is cumulative and contains the direct production and the contribution from the decay of $^{189}$Pt ($T_{1/2}$ = 10.87 h) (Fig. 19). No previous experimental data were found in the literature. The agreement with TENDL is good below 42 MeV deuteron energy and the theory gives lower values above this energy.

**Table 2** Experimental cross sections of gold radioisotopes produced on natural platinum by deuteron irradiation

| Energy E + ΔE (MeV) | | $^{199}$Au | | $^{198}$Au | | $^{196m}$Au | | $^{196g}$Au | | $^{195}$Au | | $^{194}$Au | | $^{193}$Au | | $^{192}$Au | | $^{191}$Au | |
|---|---|---|---|---|---|---|---|---|---|---|---|---|---|---|---|---|---|---|---|
| | | Cross section σ + Δσ (mb) | | | | | | | | | | | | | | | | | |
| 49.2 | 0.3 | 4.7 | 0.7 | | | 7.4 | 1.0 | 47.0 | 6.1 | 214.6 | 27.7 | 258.9 | 33.4 | 274.5 | 35.4 | 546.8 | 70.5 | 383.7 | 52.1 |
| 48.6 | 0.3 | 4.6 | 0.6 | | | | | 47.2 | 6.1 | 241.5 | 31.1 | 269.4 | 34.7 | 245.9 | 32.2 | 514.5 | 66.3 | 406.2 | 52.6 |
| 48.1 | 0.3 | 3.4 | 0.5 | | | 5.2 | 0.8 | 47.2 | 6.1 | 219.5 | 28.4 | 256.0 | 33.0 | 306.1 | 39.5 | 567.6 | 73.2 | 333.2 | 48.4 |
| 47.0 | 0.4 | 4.4 | 0.6 | | | | | 50.7 | 6.6 | 247.0 | 31.9 | 303.2 | 39.1 | 277.3 | 36.2 | 517.7 | 66.7 | 387.5 | 50.2 |
| 46.4 | 0.4 | 4.7 | 0.6 | | | 6.1 | 0.9 | 50.5 | 6.6 | 161.8 | 20.9 | 271.2 | 34.9 | 330.1 | 42.6 | 567.3 | 73.2 | 394.8 | 54.0 |
| 45.2 | 0.4 | 4.2 | 0.6 | | | | | 51.9 | 6.7 | 170.4 | 22.0 | 283.5 | 36.5 | 276.7 | 36.0 | 520.7 | 67.1 | 335.3 | 43.4 |
| 44.6 | 0.4 | 5.1 | 0.7 | | | 7.6 | 1.0 | 54.5 | 7.1 | 252.4 | 32.6 | 274.1 | 35.3 | 341.6 | 44.1 | 565.2 | 72.9 | 322.0 | 42.6 |
| 43.5 | 0.5 | 4.6 | 0.6 | | | | | 56.2 | 7.3 | 296.4 | 38.2 | 291.3 | 37.6 | 296.0 | 38.5 | 539.9 | 69.6 | 291.4 | 37.9 |
| 42.8 | 0.5 | 5.9 | 0.8 | | | 8.5 | 1.2 | 59.8 | 7.7 | 190.8 | 24.6 | 288.5 | 37.2 | 378.8 | 48.9 | 586.2 | 75.6 | 285.6 | 62.7 |
| 41.6 | 0.5 | | | | | 10.2 | 1.8 | | | 319.1 | 41.1 | | | 310.2 | 40.2 | 520.6 | 67.1 | 221.1 | 28.8 |
| 41.0 | 0.6 | 5.7 | 0.8 | | | | | 64.2 | 8.3 | 213.9 | 27.6 | 306.3 | 39.5 | | | | | | |
| 39.7 | 0.6 | 4.9 | 0.7 | | | 14.2 | 2.4 | 65.2 | 8.4 | 225.4 | 29.1 | 341.2 | 44.0 | 324.4 | 42.1 | 527.5 | 68.0 | 157.1 | 20.6 |
| 39.1 | 0.6 | 3.9 | 0.7 | | | 10.5 | 1.5 | 69.6 | 9.0 | 286.0 | 36.9 | 340.9 | 43.9 | 343.8 | 44.4 | 554.1 | 71.4 | 177.3 | 25.0 |
| 37.8 | 0.7 | 5.7 | 0.7 | | | 14.7 | 2.5 | 75.0 | 9.7 | 299.0 | 38.6 | 402.3 | 51.8 | 354.8 | 46.0 | 509.9 | 65.7 | 90.3 | 12.1 |
| 37.1 | 0.7 | 3.8 | 0.7 | 5.3 | 0.7 | 14.9 | 2.0 | 83.8 | 10.8 | 305.8 | 39.5 | 408.8 | 52.7 | 371.3 | 47.9 | 560.0 | 72.3 | | |
| 36.0 | 0.7 | 7.5 | 1.0 | | | 14.7 | 2.0 | 86.3 | 11.1 | 211.0 | 27.3 | 457.4 | 58.9 | 387.9 | 50.8 | 472.2 | 60.9 | 44.9 | 6.3 |
| 35.3 | 0.7 | 8.1 | 1.1 | 5.7 | 0.8 | 19.0 | 2.5 | 99.1 | 12.8 | 316.1 | 40.8 | 468.4 | 60.3 | 369.0 | 47.6 | 475.7 | 61.3 | | |
| 34.2 | 0.8 | 6.6 | 0.9 | | | 19.0 | 2.9 | 105.1 | 13.6 | 272.1 | 35.1 | 538.3 | 69.4 | 383.2 | 50.2 | 426.8 | 55.0 | 20.3 | 3.3 |
| 33.5 | 0.8 | 4.3 | 0.8 | 4.4 | 0.7 | 21.4 | 2.8 | 111.0 | 14.3 | 225.2 | 29.0 | 505.1 | 65.1 | 369.0 | 47.6 | 408.4 | 52.7 | | |
| 32.4 | 0.8 | 6.9 | 0.9 | | | 21.6 | 3.1 | 116.6 | 15.0 | 262.1 | 33.8 | 553.2 | 71.3 | 380.9 | 49.7 | 343.6 | 44.3 | | |
| 31.6 | 0.9 | 7.1 | 1.0 | 4.1 | 0.7 | 20.9 | 2.7 | 120.5 | 15.6 | 199.9 | 25.8 | 513.6 | 66.2 | 355.6 | 45.9 | 327.8 | 42.3 | | |
| 30.4 | 0.9 | 7.9 | 1.0 | | | 22.7 | 3.2 | 122.3 | 15.8 | 276.9 | 35.7 | 560.4 | 72.2 | 388.6 | 50.6 | 263.2 | 33.9 | | |
| 29.7 | 0.9 | 6.0 | 0.9 | 5.9 | 0.9 | 19.4 | 2.5 | 129.0 | 16.7 | 322.8 | 41.7 | 535.2 | 69.0 | 378.2 | 48.8 | 250.8 | 32.4 | | |
| 28.4 | 1.0 | 8.5 | 1.1 | 6.0 | 1.6 | 16.4 | 2.4 | 114.2 | 14.7 | 274.8 | 35.5 | 544.0 | 70.1 | 381.3 | 49.6 | 155.5 | 20.1 | | |
| 27.6 | 1.0 | 6.1 | 0.8 | 6.4 | 0.9 | 14.6 | 1.9 | 117.8 | 15.2 | 411.5 | 53.1 | 539.7 | 69.5 | 375.7 | 48.5 | 127.1 | 16.5 | | |
| 26.3 | 1.0 | 10.6 | 1.4 | 8.4 | 2.0 | 10.7 | 1.5 | 101.4 | 13.1 | 554.2 | 71.4 | 574.3 | 74.0 | 341.4 | 44.5 | 49.0 | 6.4 | | |
| 25.5 | 1.1 | 10.1 | 1.3 | 6.9 | 1.0 | 9.9 | 1.3 | 95.3 | 12.3 | 496.9 | 64.1 | 537.7 | 69.3 | 356.2 | 46.0 | 32.8 | 4.8 | | |
| 24.1 | 1.1 | 11.9 | 1.5 | 8.6 | 1.7 | 7.5 | 1.3 | 75.7 | 9.8 | 306.2 | 39.5 | 570.5 | 73.5 | 268.1 | 34.8 | 5.8 | 0.9 | | |
| 23.2 | 1.1 | 14.9 | 1.9 | 7.9 | 1.1 | 3.7 | 0.8 | 74.0 | 9.6 | 487.3 | 62.8 | 539.4 | 69.5 | 263.2 | 34.0 | 10.9 | 1.8 | | |
| 21.7 | 1.2 | 13.4 | 1.7 | 9.3 | 1.7 | 3.1 | 0.8 | 68.8 | 8.9 | 491.2 | 63.3 | 565.3 | 72.8 | 193.8 | 25.2 | 6.3 | 1.0 | | |
| 20.8 | 1.2 | 13.0 | 1.8 | 11.8 | 1.6 | 5.6 | 0.8 | | | 432.9 | 55.8 | 433.8 | 55.9 | 192.1 | 24.8 | 15.2 | 2.4 | | |

**Table 3** Experimental cross sections of platinum and iridium radioisotopes produced on natural platinum by deuteron irradiation

| Energy E + ΔE (MeV) | | $^{197m}$Pt | | $^{197g}$Pt | | $^{195m}$Pt | | $^{191}$Pt | | $^{189}$Pt | | $^{188}$Pt | | $^{194m}$Ir | | $^{192}$Ir | | $^{190}$Ir | | $^{189}$Ir | |
|---|---|---|---|---|---|---|---|---|---|---|---|---|---|---|---|---|---|---|---|---|---|
| | | Cross section σ + Δσ (mb) | | | | | | | | | | | | | | | | | | | |
| 49.2 | 0.3 | | | 45.9 | 6.9 | 68.6 | 7.7 | 608.9 | 68.5 | | | 1.2 | 0.1 | 0.61 | 0.09 | 4.3 | 0.5 | 7.1 | 0.8 | 14.8 | 1.7 |
| 48.6 | 0.3 | 29.6 | 5.5 | | | 70.7 | 8.0 | 611.6 | 68.8 | | | | | | | | | | | | |
| 48.1 | 0.3 | | | | | 72.4 | 8.2 | 564.5 | 63.5 | 9.4 | 2.4 | | | 0.49 | 0.11 | 4.0 | 0.5 | 7.2 | 0.8 | 14.9 | 1.8 |
| 47.0 | 0.4 | 30.4 | 6.2 | | | | | 444.8 | 50.0 | | | | | 0.35 | 0.06 | 4.2 | 0.5 | 7.2 | 0.8 | 12.7 | 1.6 |
| 46.4 | 0.4 | | | | | 75.9 | 8.6 | 539.4 | 60.7 | 7.1 | 2.7 | | | 0.38 | 0.07 | 4.1 | 0.5 | 7.1 | 0.8 | 11.8 | 1.5 |
| 45.2 | 0.4 | | | | | | | 384.5 | 43.2 | | | | | 0.34 | 0.04 | 3.9 | 0.4 | 6.8 | 0.8 | 11.1 | 1.3 |
| 44.6 | 0.4 | | | 46.7 | 6.8 | 77.5 | 8.7 | 464.8 | 52.3 | | | | | 0.00 | 0.00 | 4.3 | 0.5 | 6.3 | 0.7 | 11.9 | 1.5 |
| 43.5 | 0.5 | | | | | | | 321.2 | 36.1 | | | | | 0.38 | 0.05 | 4.3 | 0.5 | 6.2 | 0.7 | 9.9 | 1.2 |
| 42.8 | 0.5 | | | 55.1 | 7.4 | 80.8 | 9.1 | 385.5 | 43.4 | | | | | 0.22 | 0.05 | 4.3 | 0.5 | 6.5 | 0.8 | 8.7 | 1.1 |
| 41.6 | 0.5 | | | | | | | | | | | | | 0.27 | 0.04 | 3.8 | 0.4 | 5.7 | 0.7 | 6.9 | 0.8 |
| 41.0 | 0.6 | | | 47.1 | 6.7 | 83.7 | 9.4 | 291.3 | 32.8 | | | | | 0.23 | 0.05 | 4.1 | 0.5 | 5.6 | 0.7 | 5.3 | 0.8 |
| 39.7 | 0.6 | | | | | | | 179.7 | 20.3 | | | | | 0.29 | 0.05 | 4.2 | 0.5 | 5.6 | 0.7 | 4.5 | 0.9 |
| 39.1 | 0.6 | | | 47.3 | 6.6 | 83.2 | 9.4 | 194.8 | 22.0 | | | | | 0.37 | 0.09 | 4.4 | 0.5 | 4.8 | 0.6 | 4.0 | 0.6 |
| 37.8 | 0.7 | 20.7 | 10.2 | | | | | 102.9 | 11.6 | | | | | 0.15 | 0.04 | 4.1 | 0.5 | 4.2 | 0.5 | 2.8 | 0.5 |
| 37.1 | 0.7 | | | 51.4 | 7.0 | 96.2 | 10.9 | 100.7 | 11.5 | | | | | 0.30 | 0.12 | 4.0 | 0.5 | 4.2 | 0.5 | 1.2 | 0.3 |
| 36.0 | 0.7 | | | | | | | 47.3 | 5.3 | | | | | 0.16 | 0.05 | 4.1 | 0.5 | 3.6 | 0.5 | | |
| 35.3 | 0.7 | | | 54.3 | 6.7 | 91.6 | 10.3 | 45.7 | 5.3 | | | | | 0.24 | 0.10 | 4.3 | 0.5 | 3.5 | 0.4 | | |
| 34.2 | 0.8 | | | | | | | 20.6 | 2.5 | | | | | 0.15 | 0.03 | 4.3 | 0.5 | 3.3 | 0.4 | | |
| 33.5 | 0.8 | | | 53.3 | 7.1 | 87.6 | 9.9 | 13.3 | 2.3 | | | | | 0.15 | 0.04 | 4.2 | 0.5 | 2.8 | 0.4 | | |
| 32.4 | 0.8 | | | | | | | 8.6 | 1.3 | | | | | 0.18 | 0.04 | 3.9 | 0.4 | 2.7 | 0.3 | | |
| 31.6 | 0.9 | | | 41.5 | 6.0 | 82.4 | 9.4 | 10.0 | 1.9 | | | | | 0.12 | 0.02 | 3.8 | 0.4 | 2.1 | 0.3 | | |
| 30.4 | 0.9 | | | | | | | 9.2 | 1.1 | | | | | 0.14 | 0.03 | 3.6 | 0.4 | 2.2 | 0.3 | | |
| 29.7 | 0.9 | | | 45.6 | 6.4 | 86.3 | 9.8 | 11.8 | 2.1 | | | | | 0.14 | 0.09 | 3.6 | 0.4 | 1.2 | 0.2 | | |
| 28.4 | 1.0 | 22.2 | 8.3 | | | | | 8.7 | 1.3 | | | | | | | 3.3 | 0.4 | 1.7 | 0.3 | | |
| 27.6 | 1.0 | | | 38.5 | 5.8 | 80.2 | 9.2 | 12.8 | 2.1 | | | | | 0.14 | 0.11 | 3.5 | 0.4 | 0.8 | 0.2 | | |
| 26.3 | 1.0 | 11.1 | 3.0 | | | | | 10.2 | 1.4 | | | | | 0.13 | 0.03 | 3.2 | 0.4 | 0.9 | 0.1 | | |
| 25.5 | 1.1 | | | 51.9 | 6.9 | 78.1 | 8.9 | 16.6 | 2.6 | | | | | 0.20 | 0.11 | 3.0 | 0.4 | 0.4 | 0.1 | | |
| 24.1 | 1.1 | 7.6 | 3.6 | | | | | 8.3 | 1.3 | | | | | 0.11 | 0.04 | 2.8 | 0.3 | 0.5 | 0.2 | | |
| 23.2 | 1.1 | | | 45.8 | 6.1 | 77.7 | 8.8 | 11.8 | 2.2 | | | | | 0.19 | 0.04 | 2.7 | 0.3 | 0.2 | 0.0 | | |
| 21.7 | 1.2 | | | | | | | 6.1 | 1.1 | | | | | 0.13 | 0.02 | 2.4 | 0.3 | 0.2 | 0.1 | | |
| 20.8 | 1.2 | | | | | 73.6 | 9.1 | | | | | | | 0.21 | 0.04 | 2.1 | 0.2 | 0.2 | 0.1 | | |

## 4. Application for Thin Layer Activation (TLA)

Reviewing the produced radioisotopes (Table 1) regarding their half-life, gamma energy an intensity as well as cross sections (yields) one can select the proper isotope for wear (or corrosion and erosion) measurements by using thin layer activation [19-21]. The radioisotope $^{195g}$Au has proper half-life and cross section, but the gamma-energies are low with very low intensities. Another possible candidate for TLA is the $^{196g}$Au with 6.17 days half-life and at least one 355.7 keV 87% proper gamma-line. Its cross section is also reasonable high and it has even two maxima, which makes homogenous activity distribution [22] in the upper surface layer. The first maximum is even within the energy range of the medium energy accelerators. Iridium cross sections do not show proper maxima for homogenous activation and the values are two low for economic industrial application. In Fig. 20, as an example the different activity distributions of $^{196g}$Au are presented. By exploiting the first maximum in order to produce homogenous activity distribution in the surface layer the optimum bombarding energy is 14 MeV (see Fig. 20). By perpendicular irradiation the homogeneity range (within 1%) is 16.2 μm by 134 μm total penetration depth. If the irradiation is performed under 15$^{o}$ angle, the homogeneity range drops to 4.2 μm by 35 μm total penetration depth, and in this case the specific activity is much higher. It is also possible to exploit the higher energy maximum, if the particular task requires deeper activation. In this case the optimum bombarding energy is 30.9 MeV. The perpendicular irradiation produces less specific activity, but the homogeneity range is much larger, 51 μm by 604 μm total penetration depth. A disadvantage is in this case that a large amount of activity is deposited in the deep layers of the surface. This activity might cause radiation safety problems and useless for the wear measurement by TLA. In this case the 15$^{o}$ irradiation produces 13.2 μm homogeneous distribution by 156 μm total penetration depth. It is a very similar homogeneity range as by the lower energy perpendicular activation with the advantage of the larger specific activity and the disadvantage of the unused buried activity. According to Fig. 20 the TLA tool can be tuned in a wide range of depth according to the requirements of the particular wear measurement task.

## 5. Summary and conclusion

We report experimental cross sections for production of $^{nat}Pt(d,x)^{191,192,193,194,195,196m2,196g,198g,199}Au$, $^{nat}Pt(d,x)^{188,189,191,195m,197g,197m}Pt$ and $^{nat}Pt(d,x)^{189,190,192,194m2}Ir$ reactions in the 20-50 MeV energy range. The new data are the first experimental data sets above 40 MeV for all isotopes. For production of $^{197m,189,188}Pt$ and $^{194m2,190,189}Ir$ no earlier experimental data were found. A partial or good agreement was found with the results of our previous studies in the overlapping energy regions. The comparison with the theoretical predictions as recently published TENDL-2014 and 2015 libraries shows some modification in the codes, but the agreement is still moderate in some cases, especially for isomeric cross sections. No remarkable improvement can be observed between the recent and previous versions of the TENDL libraries. Knowledge of the excitation functions is significant for different applications, among others for production of medically relevant $^{198,199}Au$ radioisotopes, for accelerator technology and targetry. The applicability of the produced radioisotopes for wear measurements tasks has been demonstrated through the example of $^{196g}Au$.

## Acknowledgments

This work was performed in the frame of the HAS-FWO Vlaanderen (Hungary-Belgium) project. The authors acknowledge the support of the research project and of the respective institutions. We thank to Cyclotron Laboratory of the Université Catholique in Louvain la Neuve (LLN) providing the beam time and the crew of the LLN Cyclone 90 cyclotron for performing the irradiations.

**Figures**

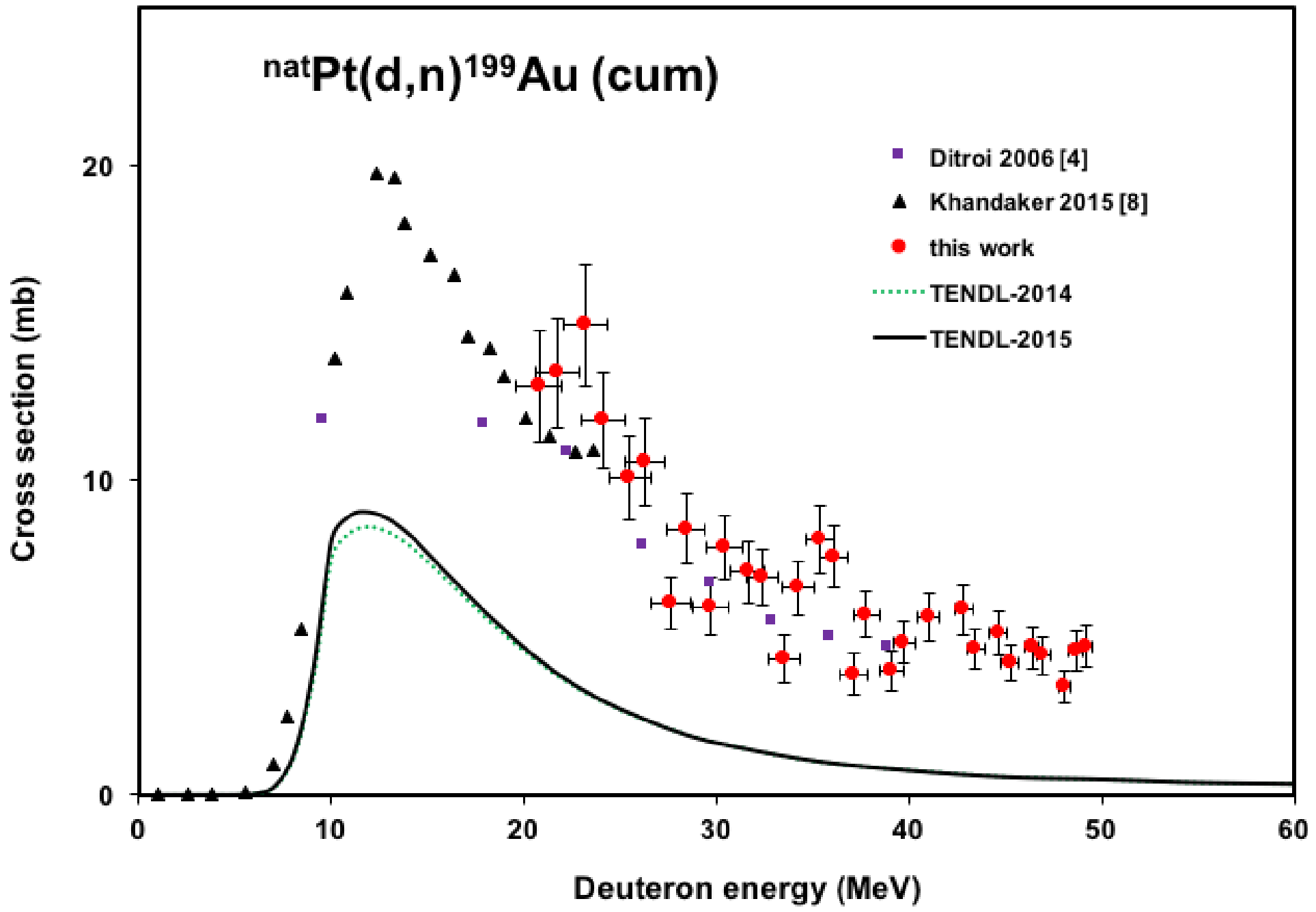

Fig. 1. Excitation function of the $^{nat}Pt(d,n)^{199}Au$ reaction

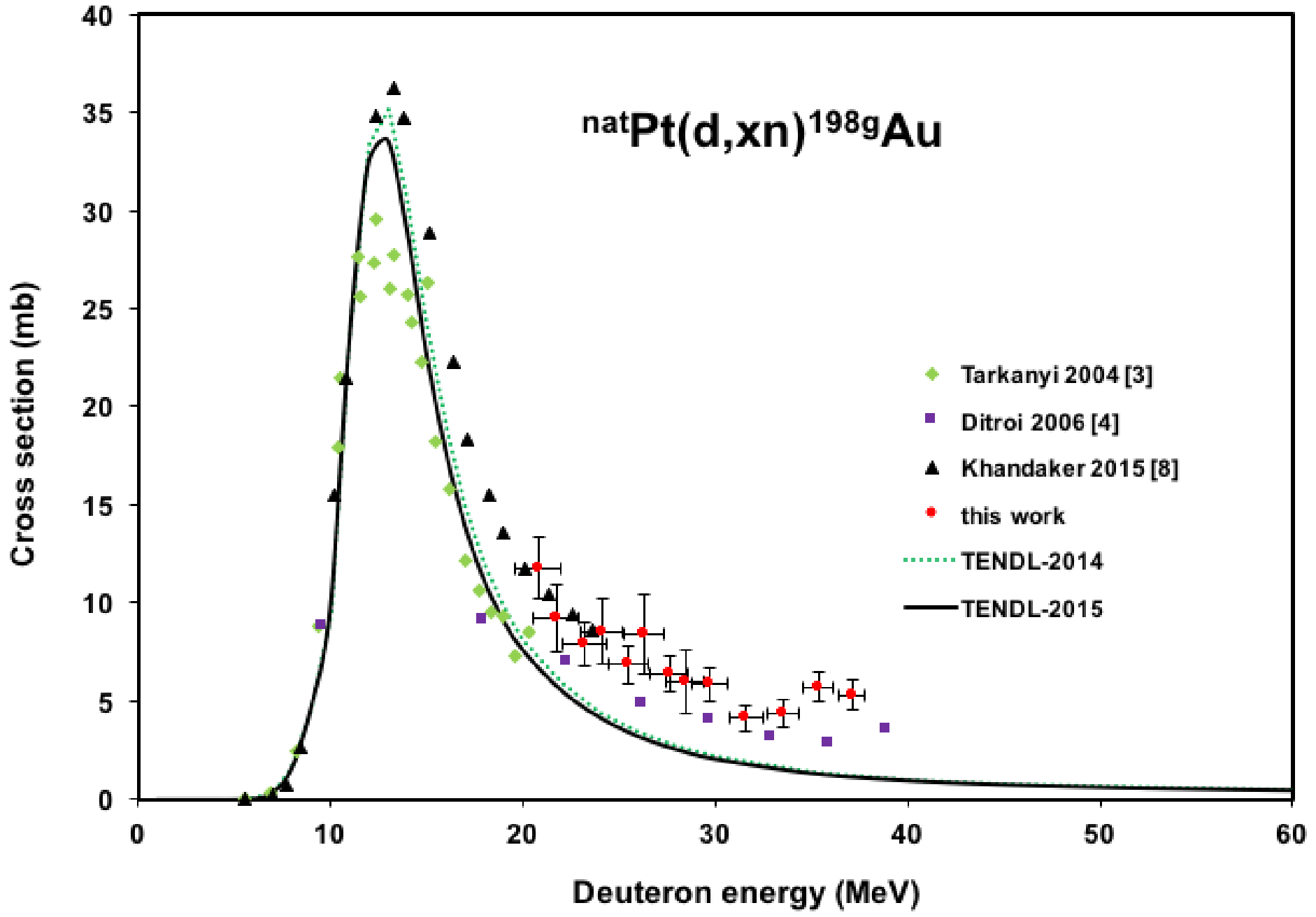

Fig. 2. Excitation function of the $^{nat}Pt(d,xn)^{198g}Au$ reaction

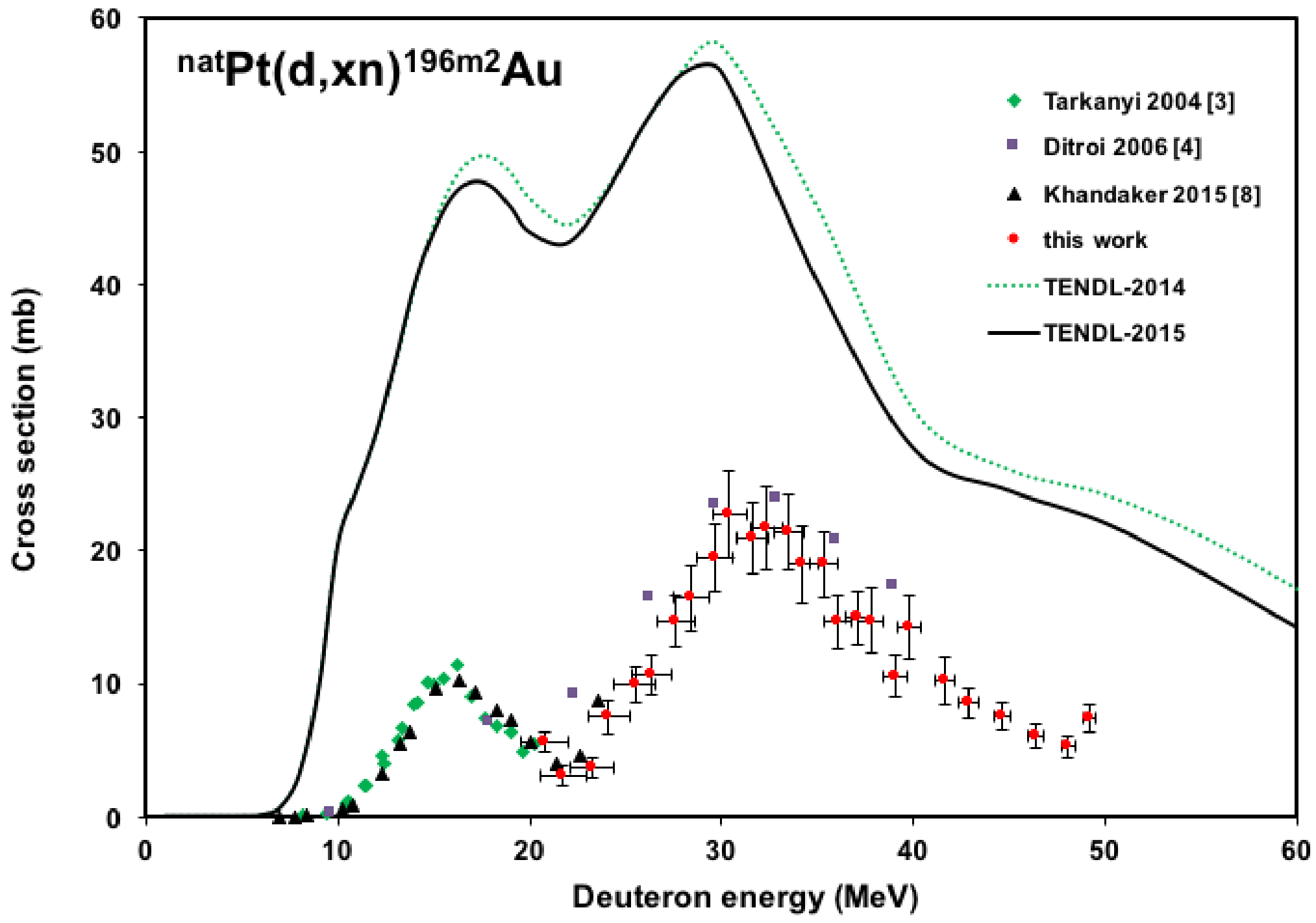

Fig. 3. Excitation function of the ${}^{nat}Pt(d,xn){}^{196m2}Au$ reaction

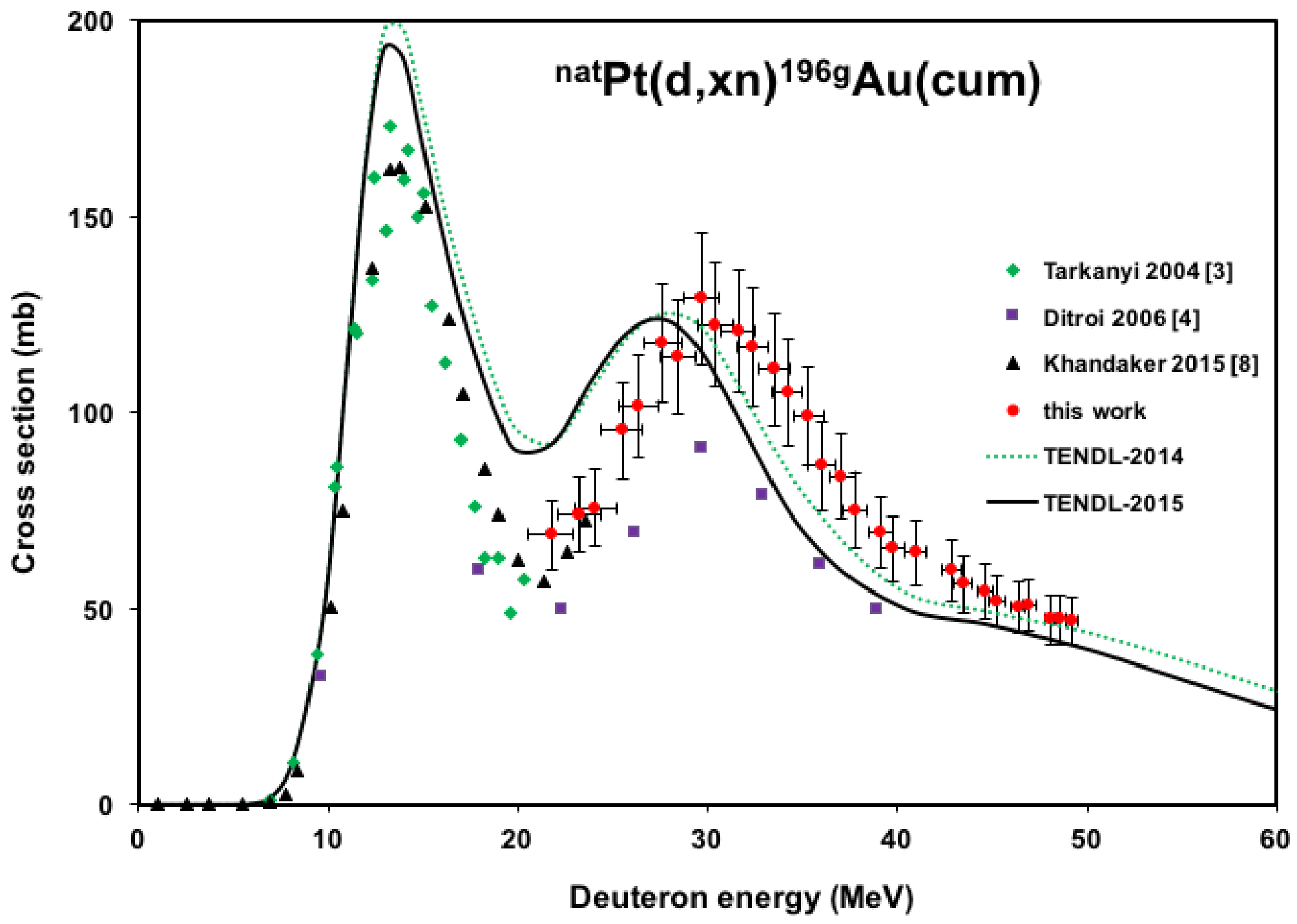

Fig. 4. Excitation function of the ${}^{nat}Pt(d,xn){}^{196g}Au$ (cum) reaction

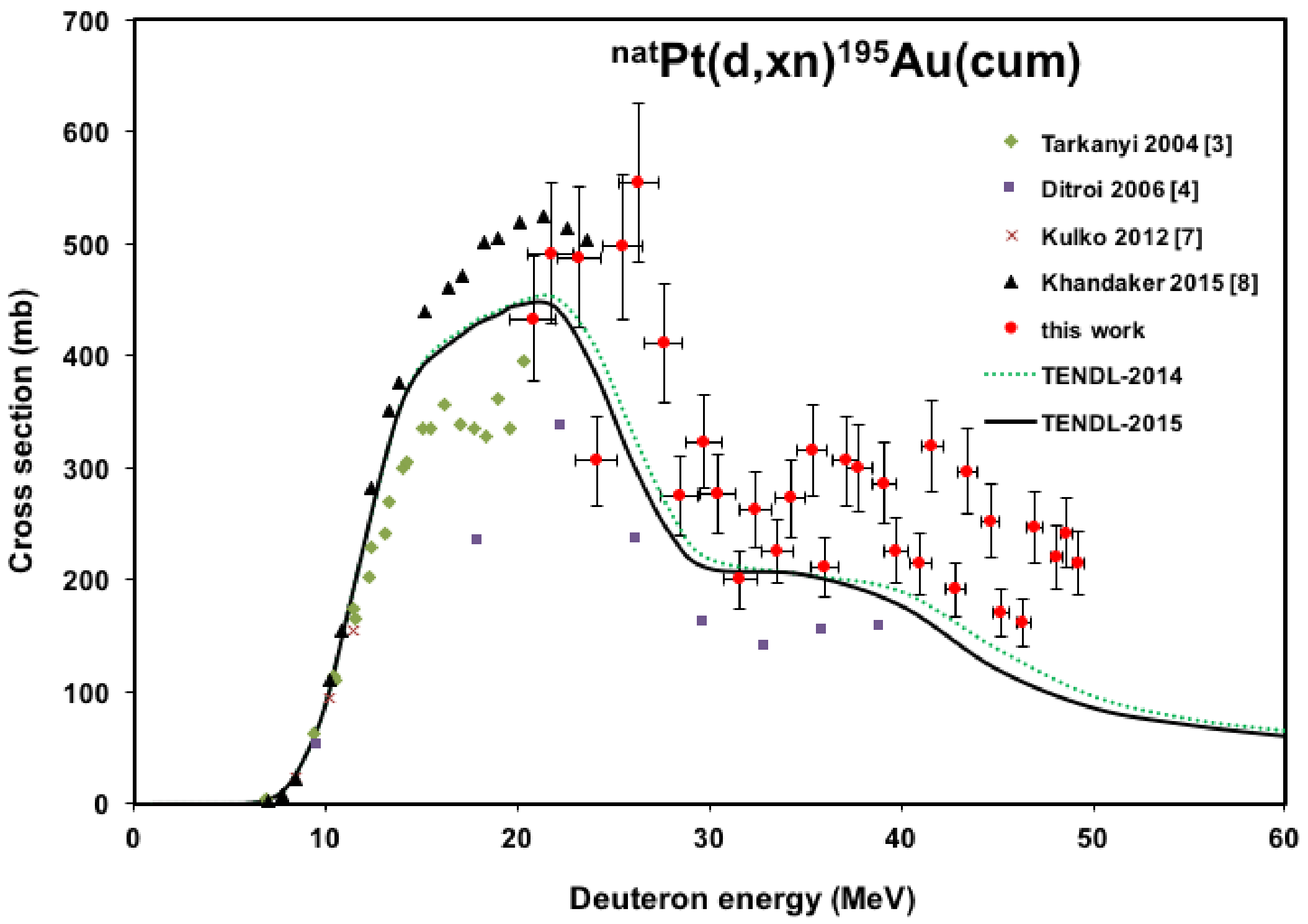

Fig. 5. Excitation function of the $^{nat}Pt(d,xn)^{195}Au$ (cum) reaction

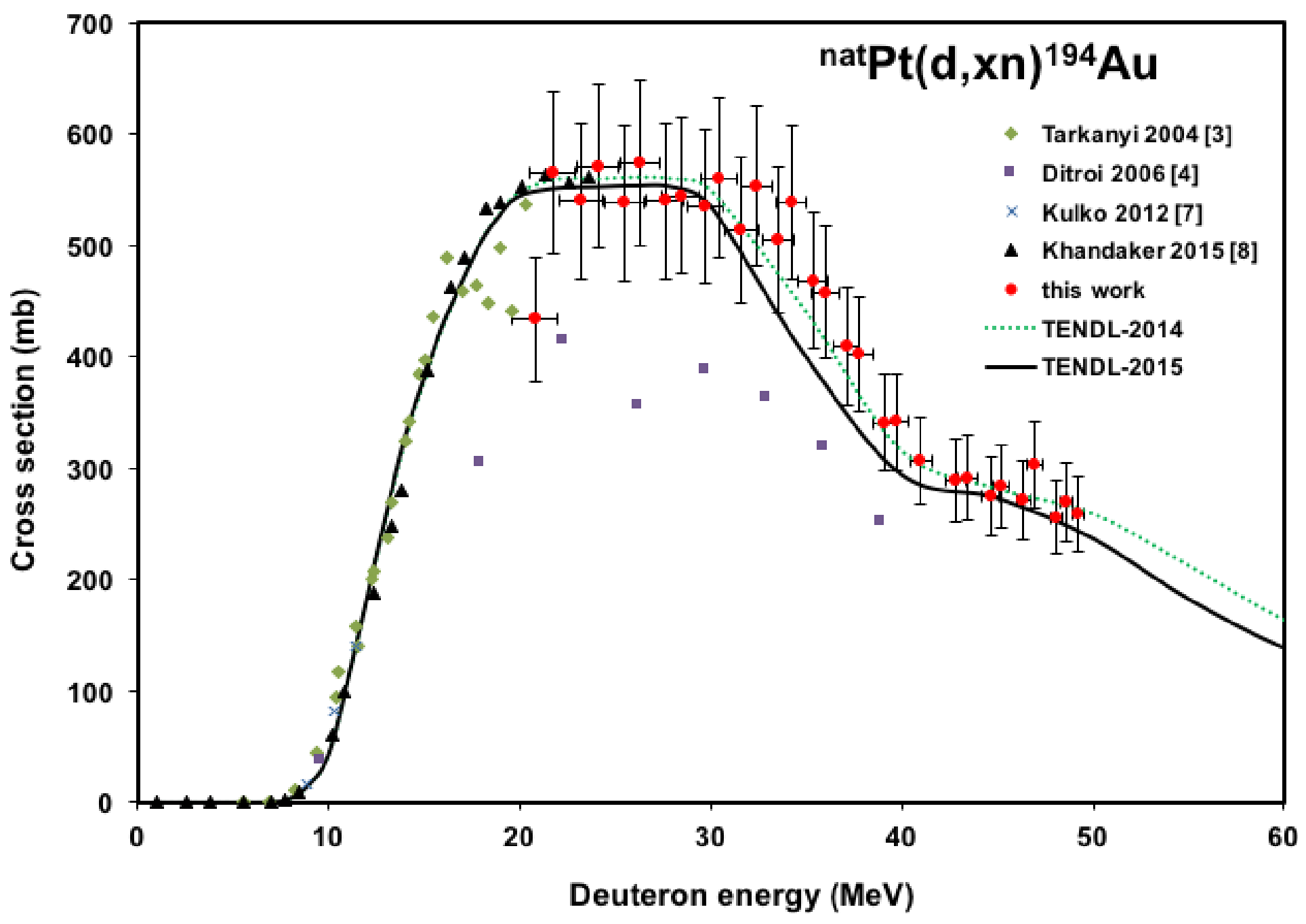

Fig. 6. Excitation function of the $^{nat}Pt(d,xn)^{194}Au$ reaction

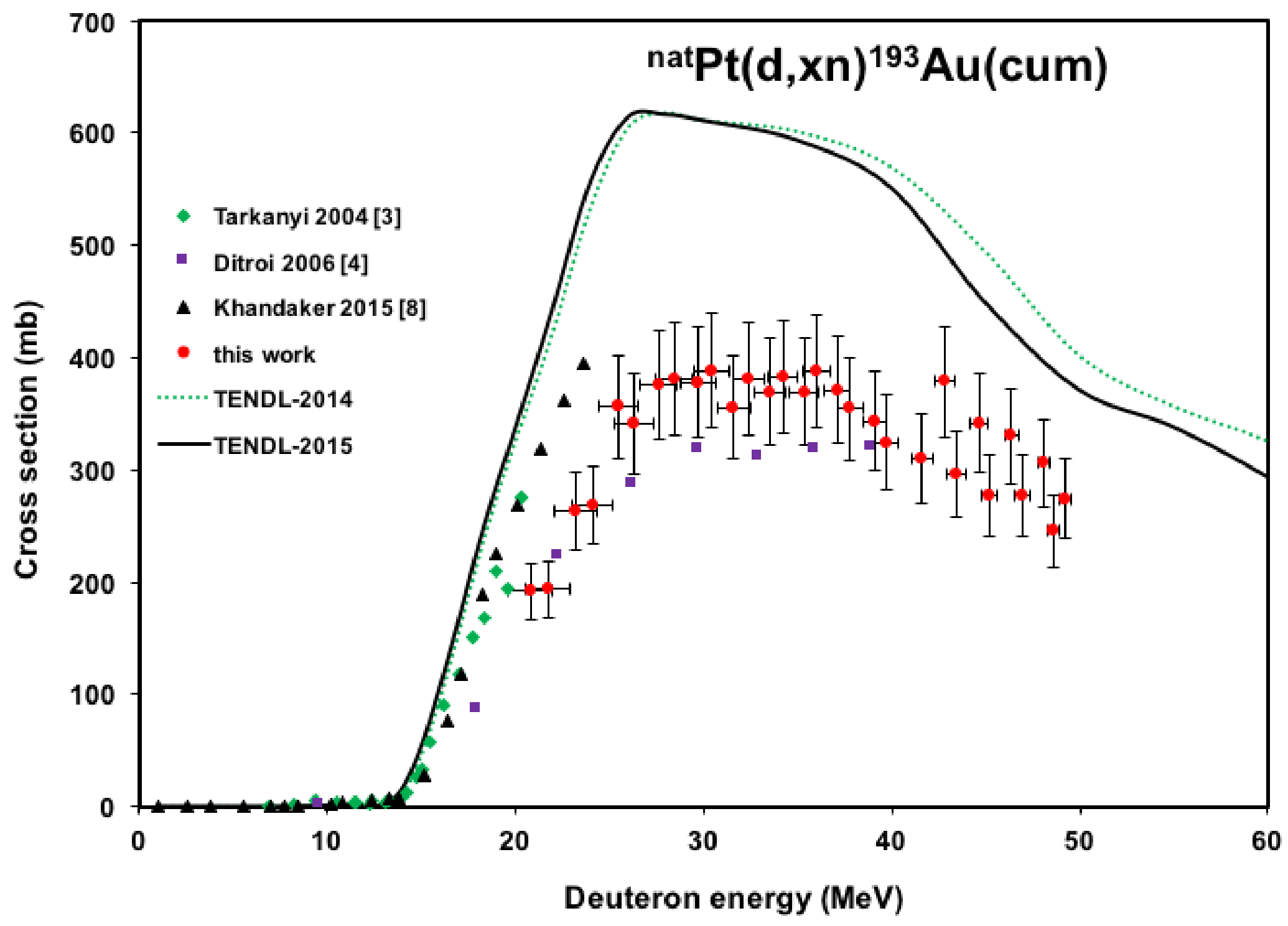

Fig. 7. Excitation function of the $^{nat}Pt(d,xn)^{193}Au$ reaction

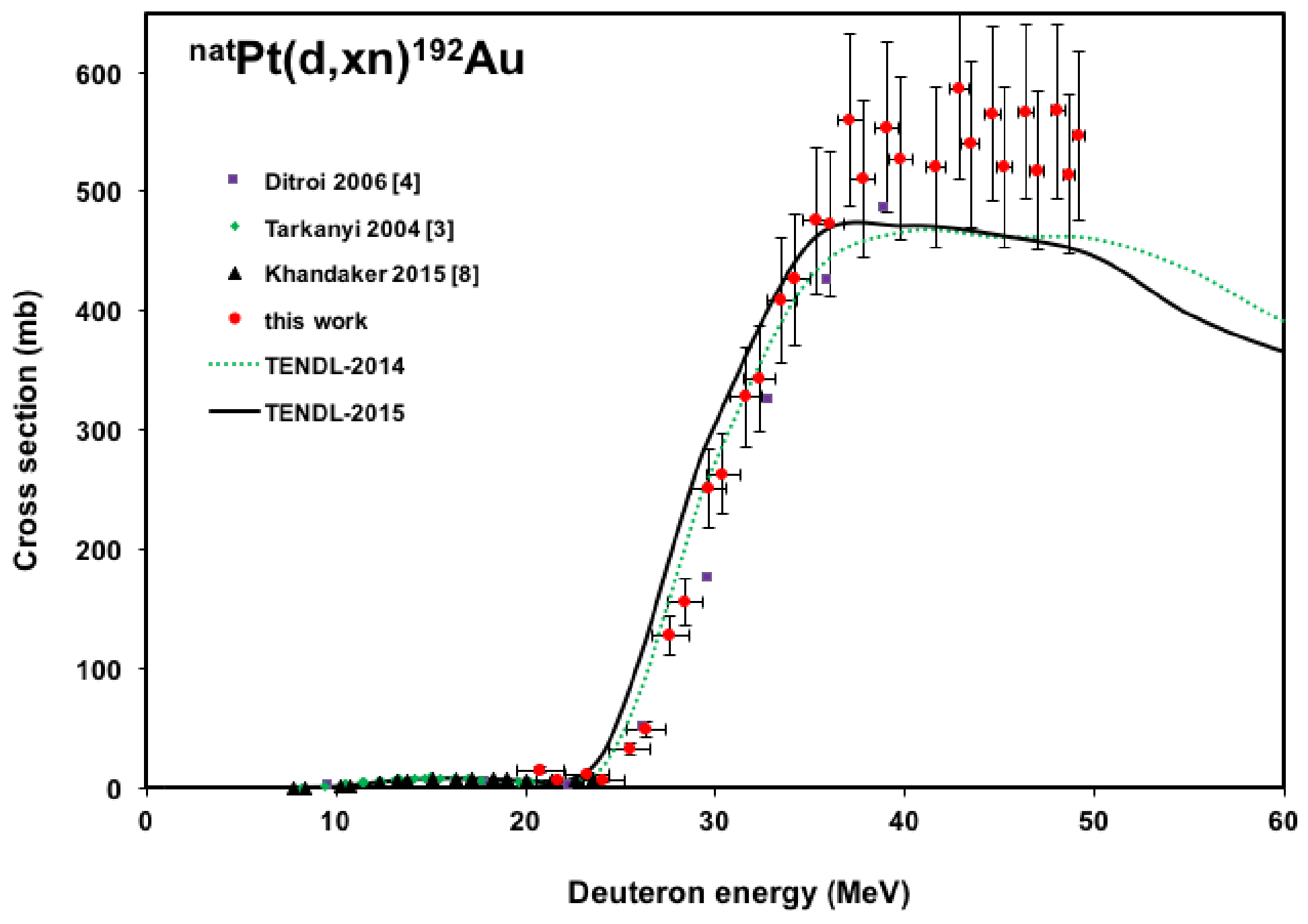

Fig. 8. Excitation function of the $^{nat}Pt(d,xn)^{192}Au$ reaction

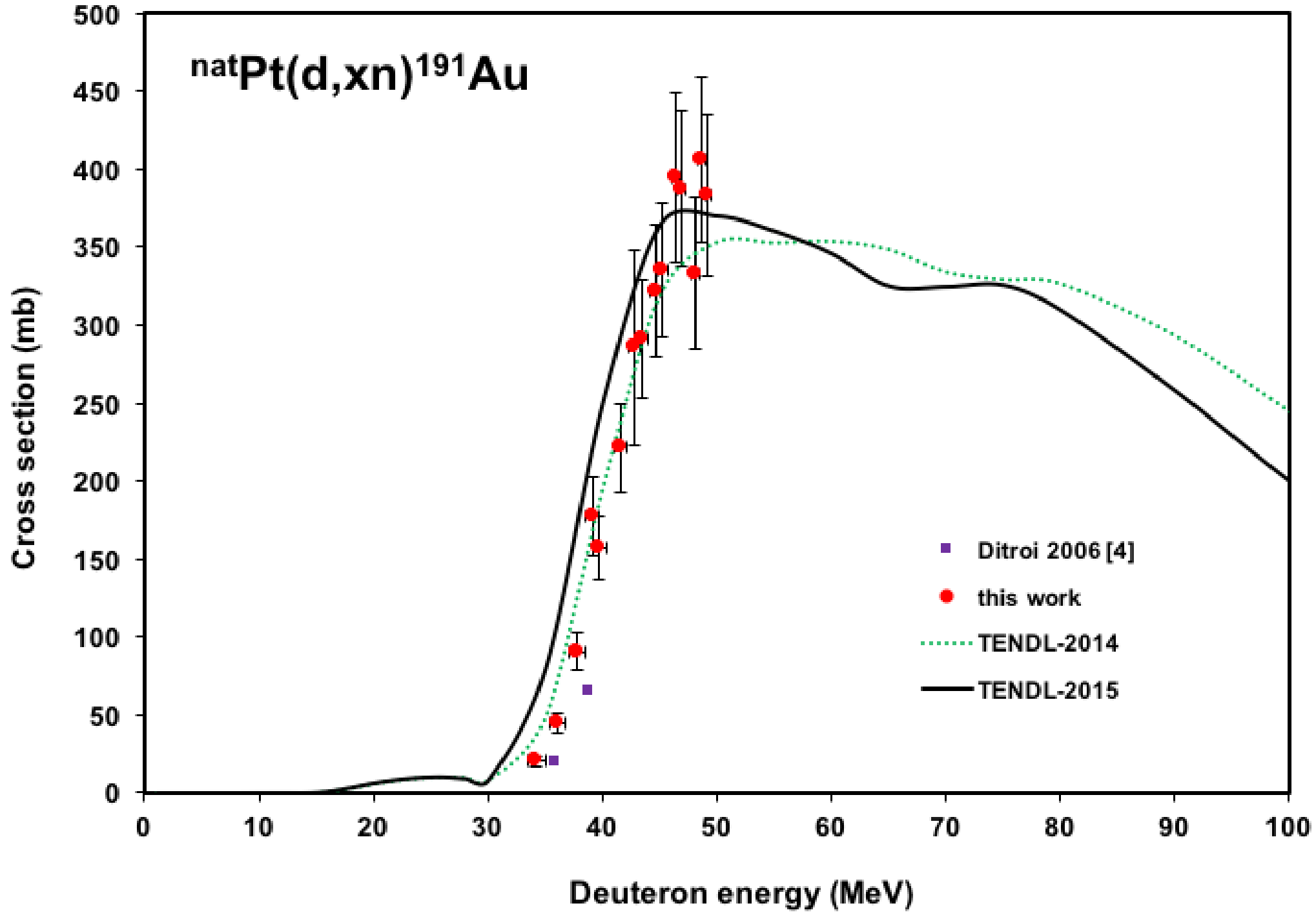

Fig. 9. Excitation function of the ${}^{nat}Pt(d,xn)^{191}Au$ reaction

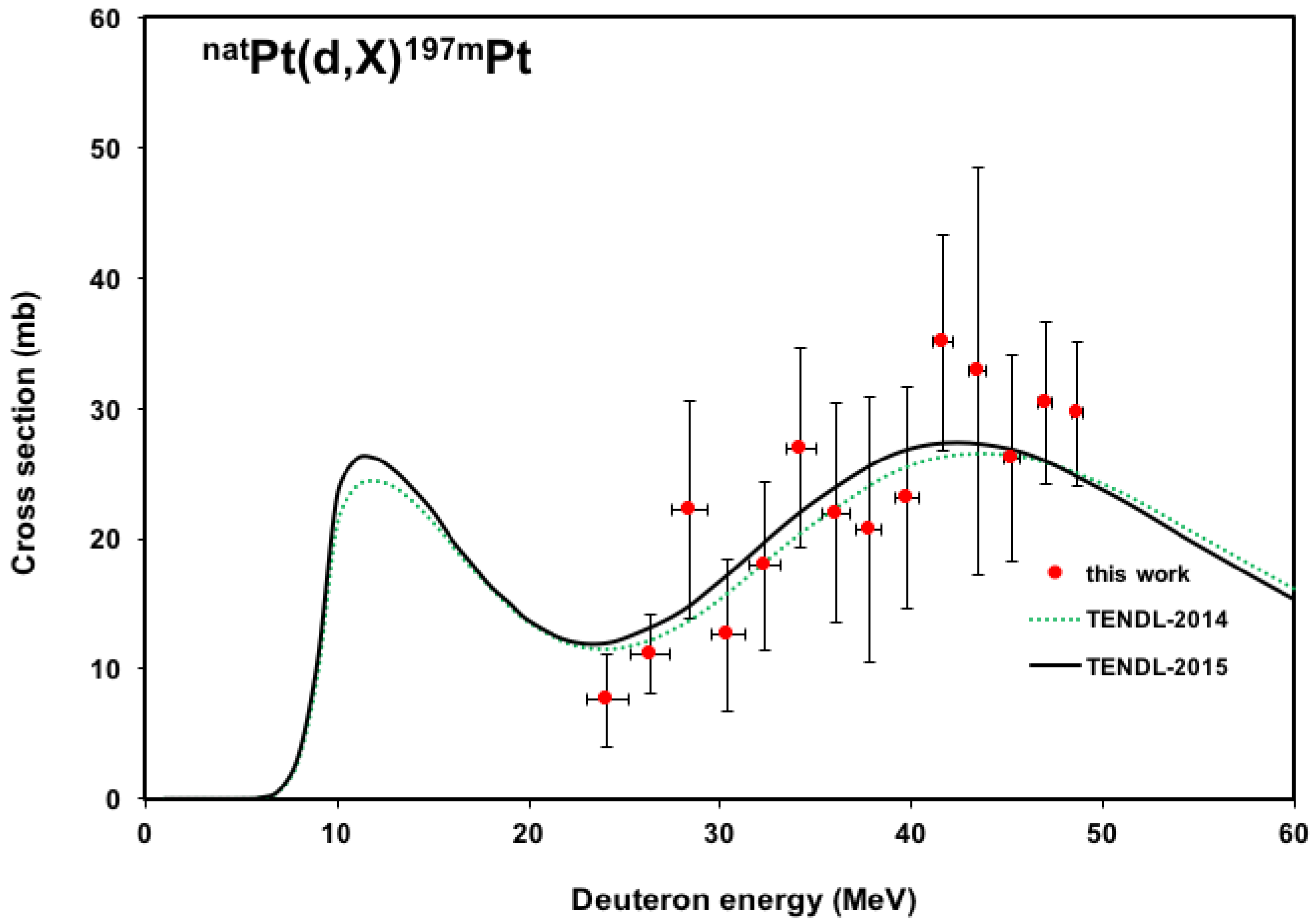

Fig. 10. Excitation function of the ${}^{nat}Pt(d,X)^{197m}Pt$ reaction

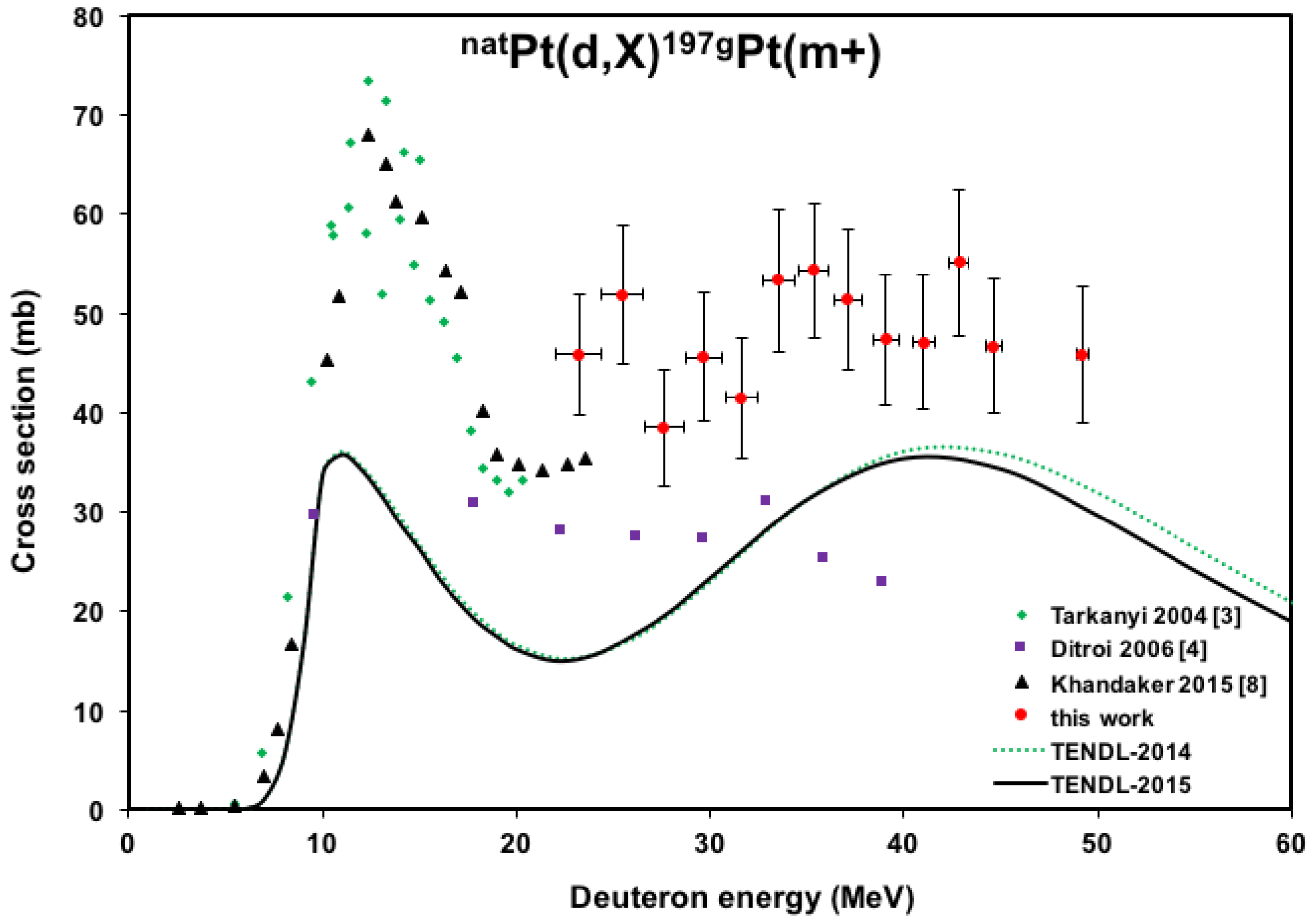

Fig. 11. Excitation function of the $^{nat}Pt(d,X)^{197g}Pt$ reaction

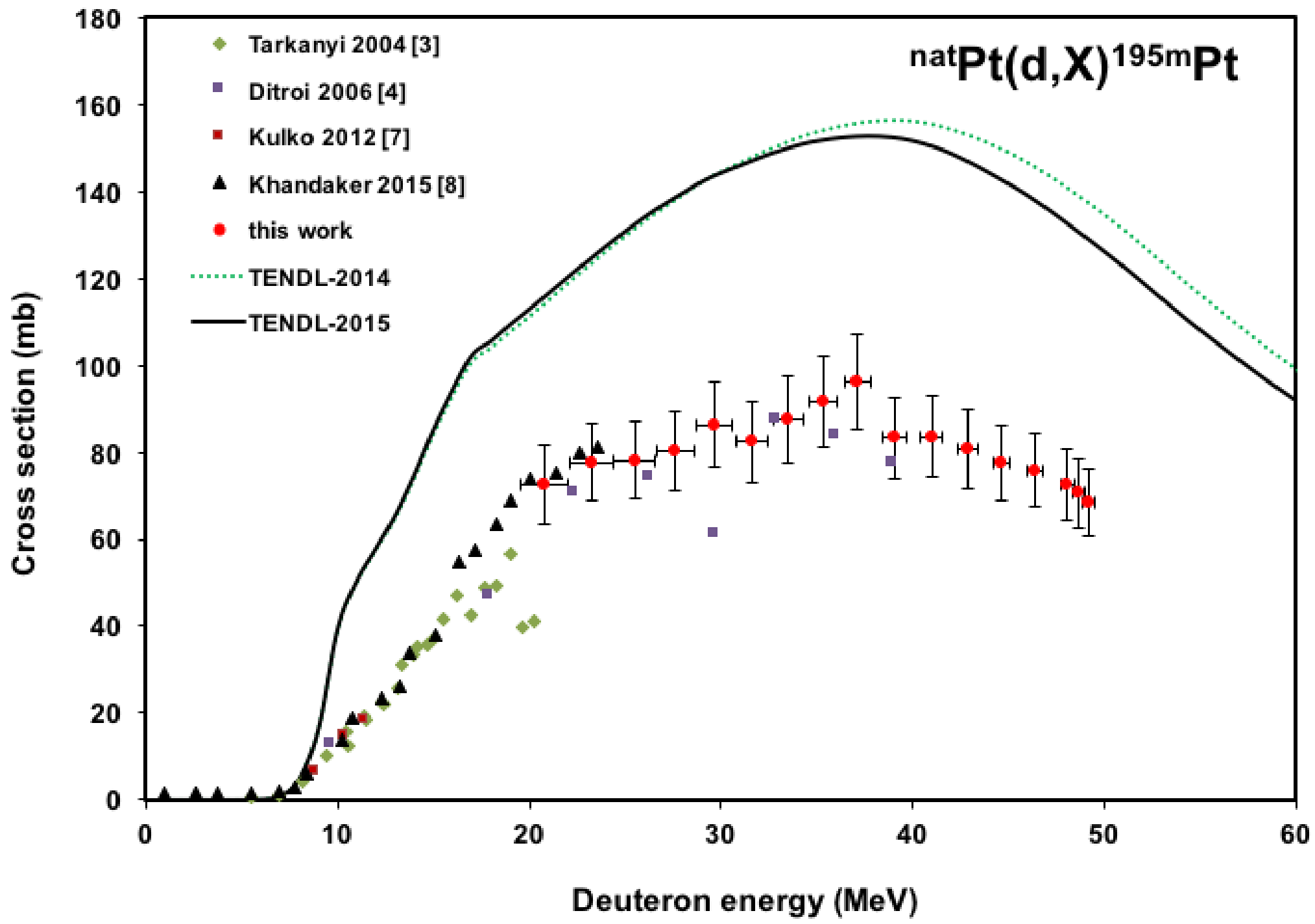

Fig. 12. Excitation function of the $^{nat}Pt(d,X)^{195m}Pt$ reaction

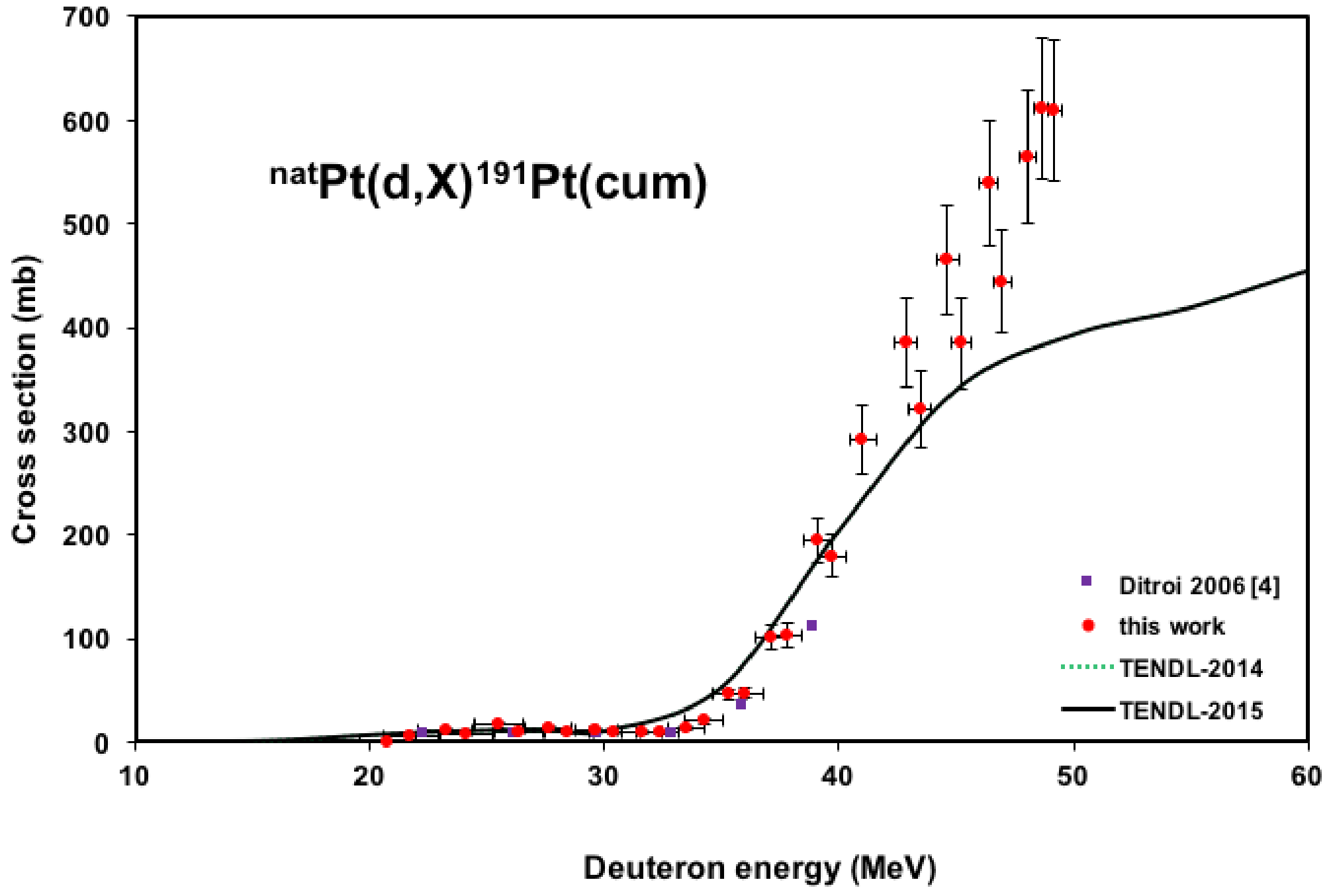

Fig. 13. Excitation function of the $^{nat}Pt(d,X)^{191}Pt$ reaction

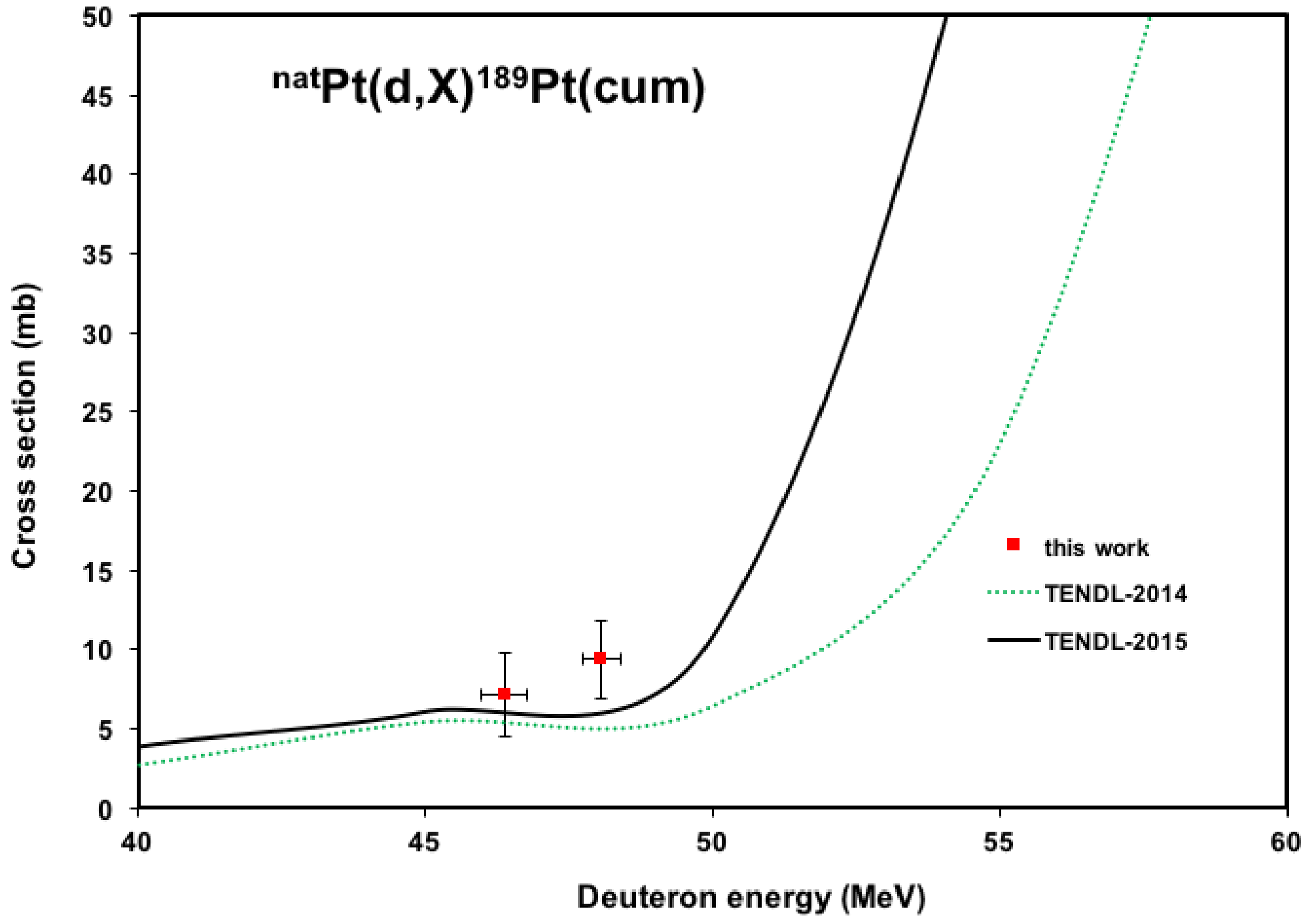

Fig. 14. Excitation function of the $^{nat}Pt(d,X)^{189}Pt$ reaction

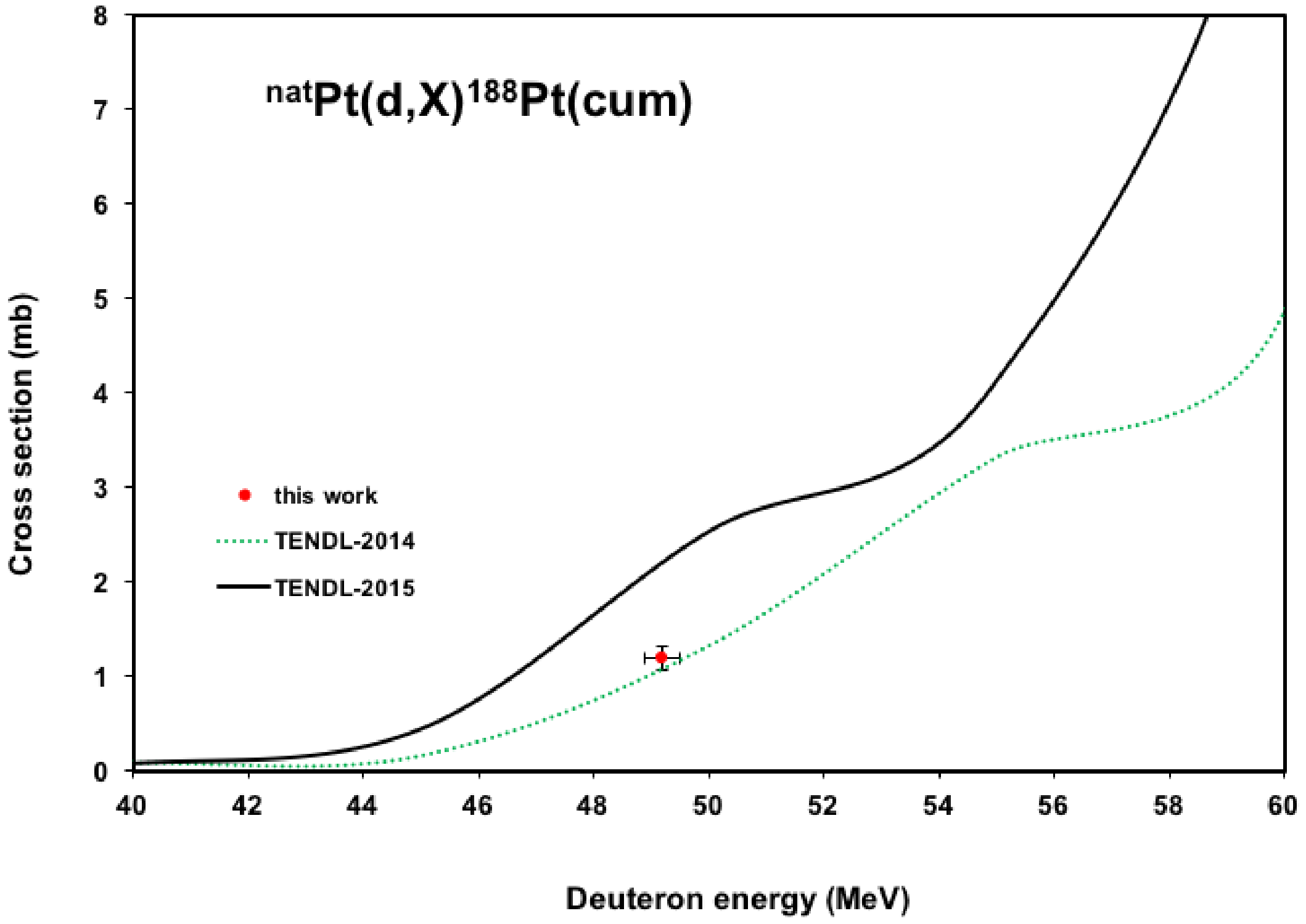

Fig. 15. Excitation function of the $^{nat}Pt(d,X)^{188}Pt$ reaction

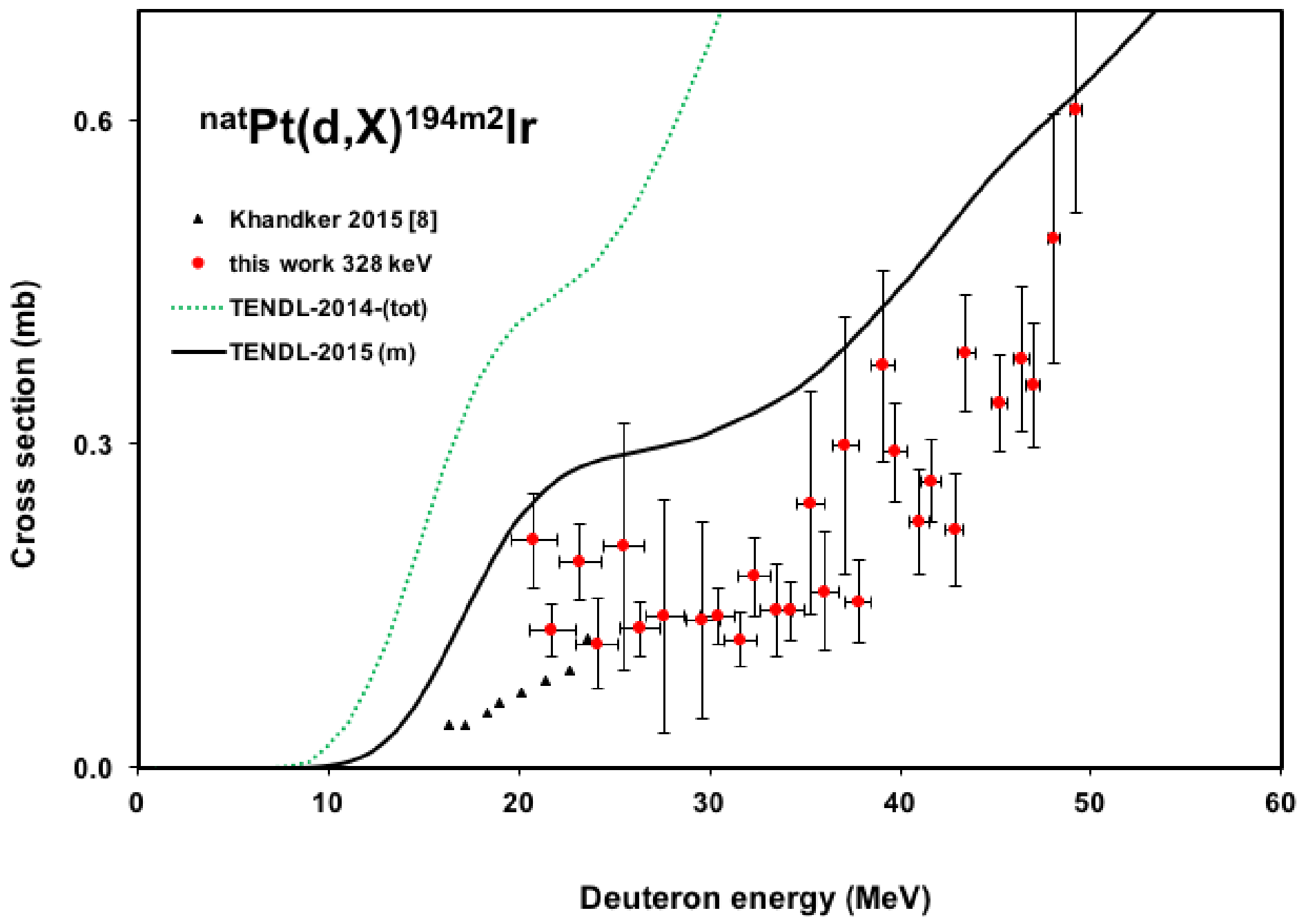

Fig. 16. Excitation function of the $^{nat}Pt(d,X)^{194m2}Ir$ reaction

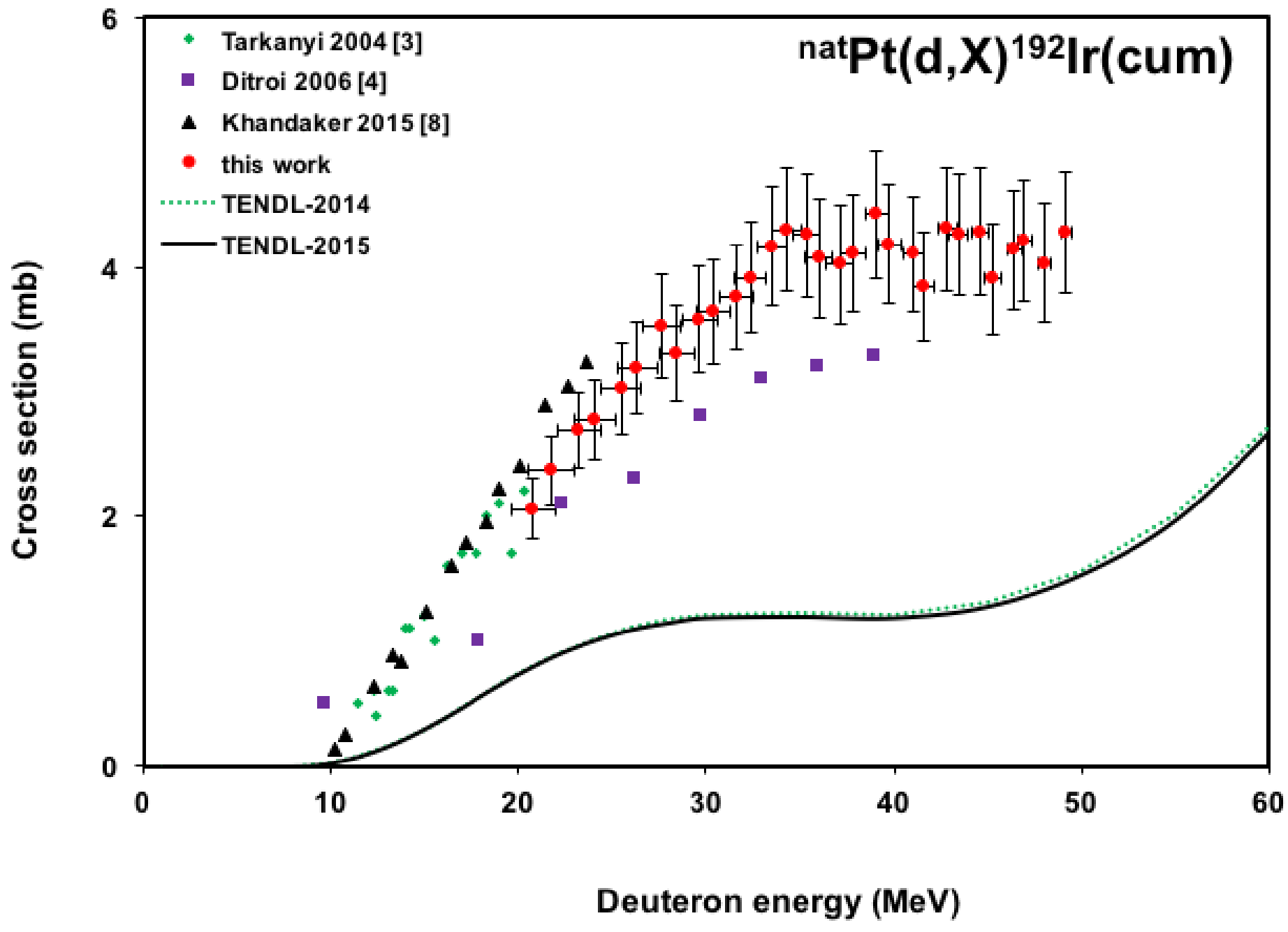

Fig. 17. Excitation function of the $^{nat}Pt(d,X)^{192}Ir$ reaction

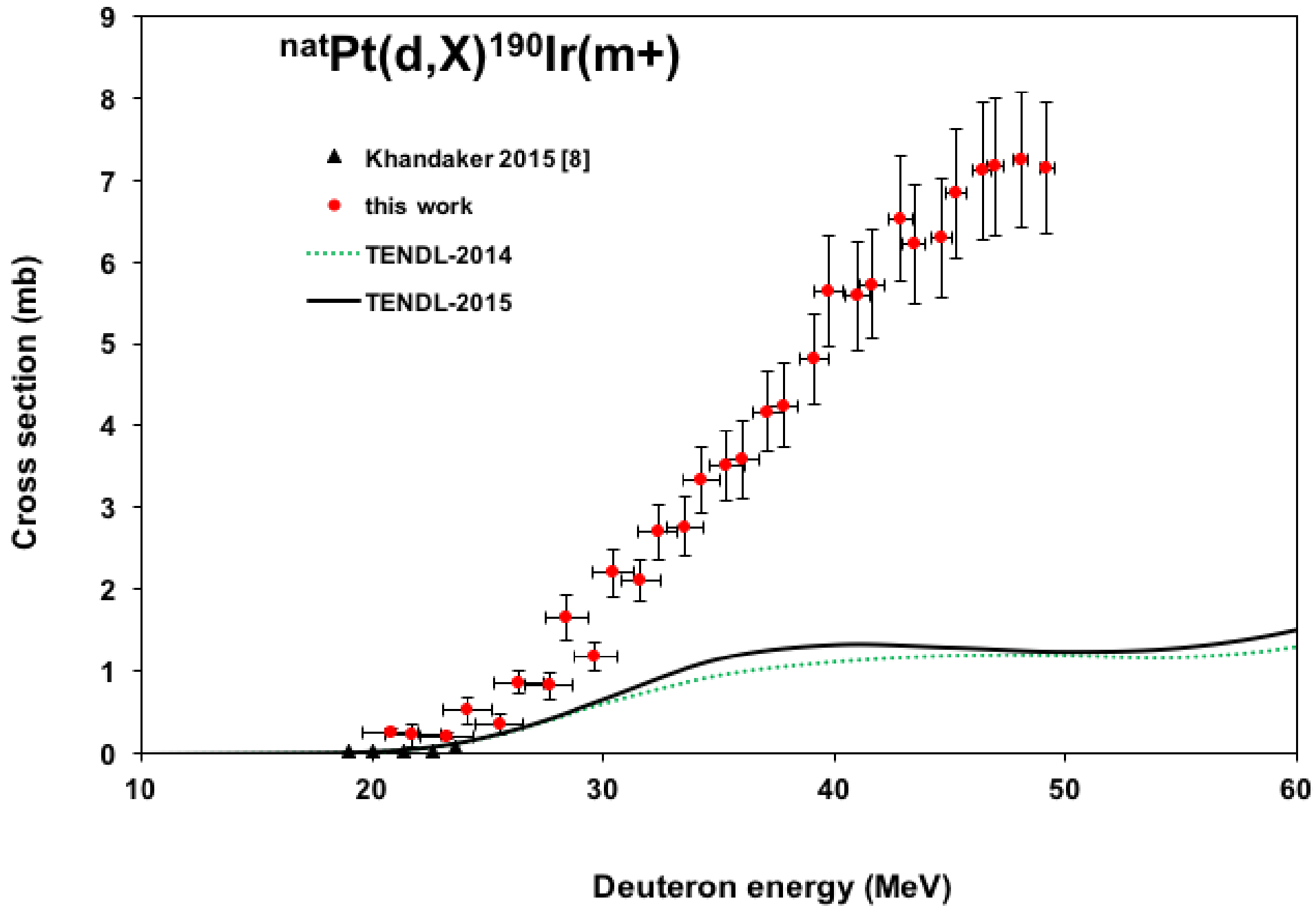

Fig. 18. Excitation function of the $^{nat}Pt(d,X)^{190}Ir$ reaction

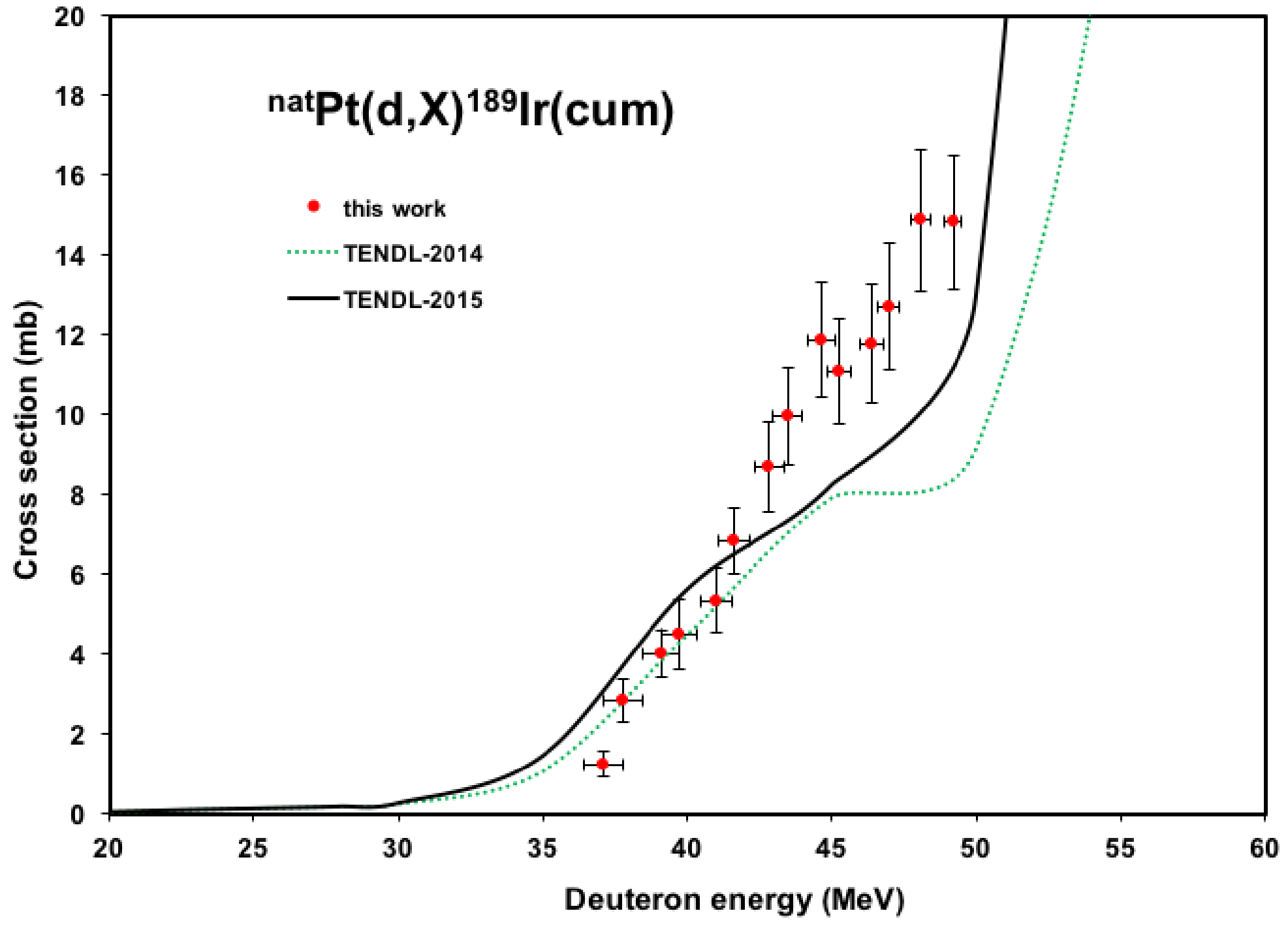

Fig. 19. Excitation function of the $^{nat}Pt(d,X)^{189}Ir$ reaction

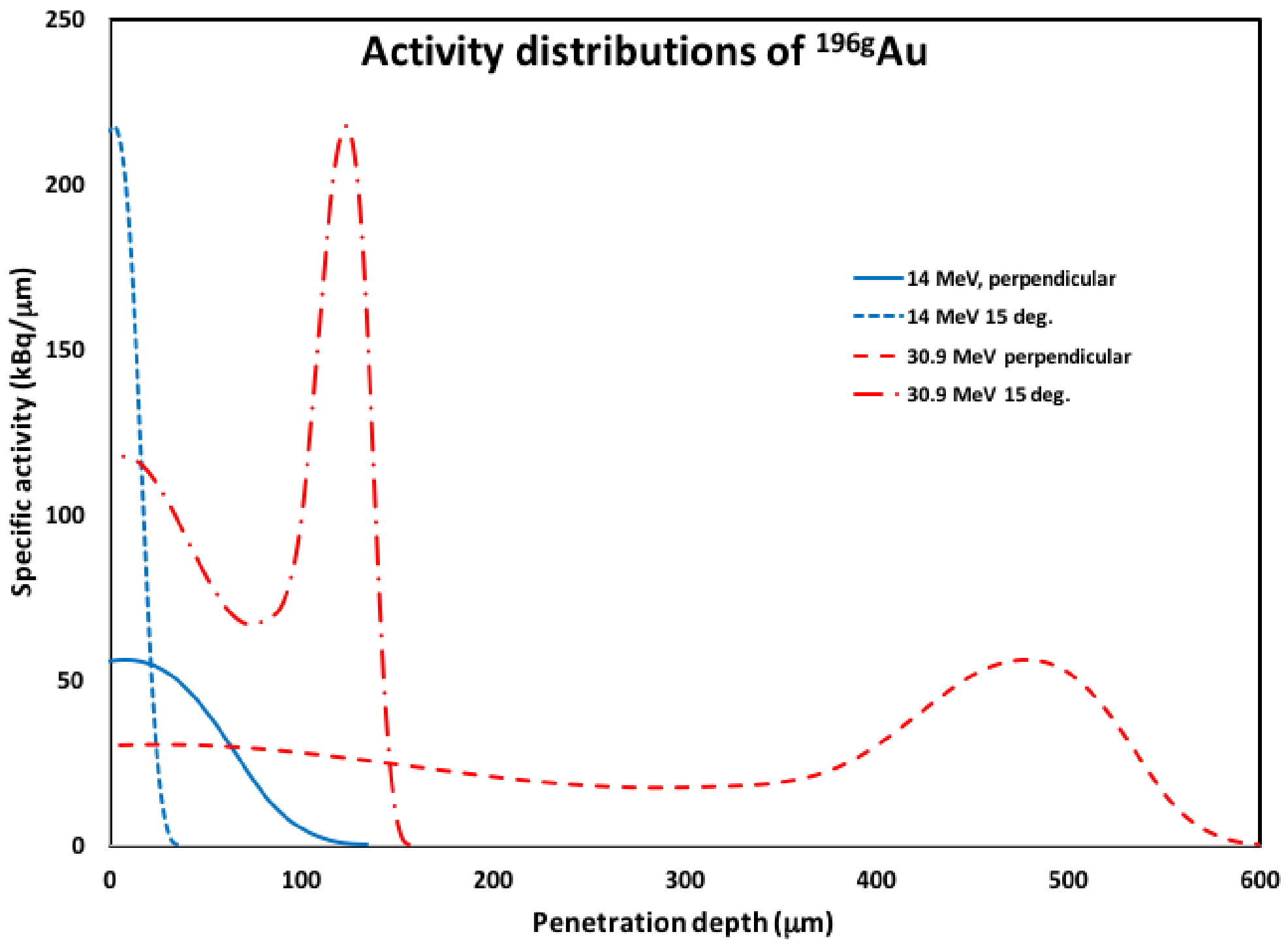

Fig. 20. Activity versus depth distributions of the $^{196g}Au$ radioisotope under different irradiation conditions (see in text)

# References

[1] F. Tárkányi, A. Hermanne, S. Takács, Y.N. Shubin, A.I. Dityuk, Cross sections for production of the therapeutic radioisotopes Au-198 and Au-199 in proton and deuteron induced reactions on Pt-198, Radiochim. Acta, 92 (2004) 223-228.

[2] F. Tárkányi, S. Takács, F. Ditrói, G. Csikai, A. Hermanne, M.S. Uddin, M. Hagiwara, M. Baba, Y.N. Shubin, A.I. Dityuk, Measurement of activation cross sections of the proton, deuteron, and alpha particle-induced nuclear reactions on platinum, in: R.C. Haight, P. Talou, T. Kawano (Eds.) International Conference on Nuclear Data for Science and Technology, AIP, Santa Fe, USA, 2004, pp. 1015-1018.

[3] F. Tárkányi, S. Takács, F. Ditrói, A. Hermanne, Y.N. Shubin, A.I. Dityuk, Activation cross-sections of deuteron induced reactions on platinum, Nucl. Instrum. Methods Phys. Res., Sect. B, 226 (2004) 490-498.

[4] F. Ditrói, F. Tárkányi, J. Csikai, M.S. Uddin, M. Hagiwara, M. Baba, Y.N. Shubin, S.F. Kovalev, Excitation functions of long lived products in deuteron induced nuclear reactions on platinum up to 40 MeV, Nucl. Instrum. Methods Phys. Res., Sect. B, 243 (2006) 20-27.

[5] F. Tárkányi, F. Ditrói, S. Takács, J. Csikai, A. Hermanne, M.S. Uddin, M. Hagiwara, M. Baba, Y.N. Shubin, A.I. Dityuk, Activation cross-sections of light ion induced nuclear reactions on platinum: proton induced reactions, Nucl. Instrum. Methods Phys. Res., Sect. B, 226 (2004) 473-489.

[6] P.P. Dmitriev, N.N. Krasnov, G.A. Molin, Radioactive nuclide yields for thick target at 22 MeV deuterons energy, Yadernie Konstanti 34 (1982) 38.

[7] A.A. Kulko, N.K. Skobelev, V. Kroha, Y.E. Penionzhkevich, J. Mrázek, V. Burjan, Z. Hons, E. Šimecková, S. Pisko , A. Kugler, N.A. Demekhina, Y.G. Sobolev, T.V. Chuvilskaya, A.A. Shirokova, K. Kuterbekov, Excitation functions for deuterium Induced reactions on $^{194}$Pt near the Coulomb barrier, Phys.Part. and Nucl.Lett., 9 (2012) 502.

[8] M.U. Khandaker, H. Haba, M. Murakami, N. Otuka, H. Abu Kassim, Excitation functions of deuteron-induced nuclear reactions on natural platinum up to 24 MeV, Nucl. Instrum. Methods Phys. Res., Sect. B, 362 (2015) 151-162.

[9] A. Hermanne, F. Tárkányi, S. Takács, F. Ditrói, Z. Szücs, Activation cross sections of deuteron-induced nuclear reactions on mercury up to 50 MeV, J. Radioanal. Nucl. Chem., 308 (2015) 221-239.

[10] H.H. Andersen, J.F. Ziegler, Hydrogen stopping powers and ranges in all elements. The stopping and ranges of ions in matter, Volume 3., Pergamon Press, New York, 1977.

[11] F. Tárkányi, F. Szelecsényi, S. Takács, Determination of effective bombarding energies and fluxes using improved stacked-foil technique, Acta Radiol., Suppl., 376 (1991) 72.

[12] F. Tárkányi, S. Takács, K. Gul, A. Hermanne, M.G. Mustafa, M. Nortier, P. Oblozinsky, S.M. Qaim, B. Scholten, Y.N. Shubin, Z. Youxiang, Beam monitor reactions (Chapter 4). Charged particle cross-section database for medical radioisotope production: diagnostic radioisotopes and monitor reactions. , in:  TECDOC 1211, IAEA, 2001, pp. 49.

[13] A.A. Sonzogni, NuDat 2.0: Nuclear Structure and Decay Data on the Internet, http://www.nndc.bnl.gov/nudat2, in: M.B.C.e. Robert C. Haight (ed.) , Toshihiko Kawano (ed.) , Patrick Talou (ed.) (Ed.) ND2004 - International Conference on Nuclear Data for Science and Technology, AIP Conference Proceedings, September 26 - October 1, 2004, Santa Fe, New Mexico, USA, 2005, pp. 547-577.

[14] B. Pritychenko, A. Sonzogni, Q-value calculator, in, NNDC, Brookhaven National Laboratory, 2003.

[15] International-Bureau-of-Weights-and-Measures, Guide to the expression of uncertainty in measurement, 1st ed., International Organization for Standardization, Genève, Switzerland, 1993.

[16] A.J. Koning, D. Rochman, S. van der Marck, J. Kopecky, J.C. Sublet, S. Pomp, H. Sjostrand, R. Forrest, E. Bauge, H. Henriksson, O. Cabellos, S. Goriely, J. Leppanen, H. Leeb, A. Plompen, R. Mills, TENDL-2014: TALYS-based evaluated nuclear data library, in, www.talys.eu/tendl2014.html, 2014.

[17] A.J. Koning, D. Rochman, J. Kopecky, J.C. Sublet, E. Bauge, S. Hilaire, P. Romain, B. Morillon, H. Duarte, S. van der Marck, S. Pomp, H. Sjostrand, R. Forrest, H. Henriksson, O. Cabellos, G. S., J. Leppanen, H. Leeb, A. Plompen, R. Mills, TENDL-2015: TALYS-based evaluated nuclear data library,, in, https://tendl.web.psi.ch/tendl_2015/tendl2015.html, 2015.

[18] A.J. Koning, D. Rochman, Modern Nuclear Data Evaluation with the TALYS Code System, Nucl. Data Sheets, 113 (2012) 2841.

[19] F. Ditrói, P. Fehsenfeld, A.S. Khanna, I. Konstantinov, I. Majhunka, P.M. Racolta, T. Sauvage, J. Thereska, The thin layer activation method and its applications in industry, in:  IAEA TECDOC-924, Vienna, 1997.

[20] F. Ditrói, S. Takács, F. Tárkányi, E. Corniani, R.W. Smith, M. Jech, T. Wopelka, Sub-micron wear measurement using activities under the free handling limit, J. Radioanal. Nucl. Chem., 292 (2012) 1147-1152.

[21] F. Ditrói, S. Takács, F. Tárkányi, M. Reichel, M. Scherge, A. Gerve, Thin layer activation of large areas for wear study, Wear, 261 (2006) 1397-1400.

[22] F. Ditrói, F. Tárkányi, S. Takács, A. Hermanne, A.V. Ignatyuk, Activation cross sections of deuteron induced reactions on niobium in the 30–50 MeV energy range, Nucl. Instrum. Methods Phys. Res., Sect. B, 373 (2016) 17-27.